\documentclass[aip,reprint,apl]{revtex4-1}
\usepackage{amsmath} 
\usepackage{graphicx,color}
\usepackage[dvipsnames]{xcolor}
\usepackage{hyperref}
\hypersetup{
    pdftoolbar=true,
    pdfmenubar=true,
    bookmarksopen=true,
    bookmarksnumbered=true,
    breaklinks=true,
    linktocpage,
    colorlinks=true,
    linkcolor={blue!95!black},
    citecolor={teal!70!black},
    urlcolor={blue!95!black},
}

\begin{document}


\title{Unified Microscopic Theory of Stress Relaxation, Structural Evolution, and Memory Effects in Dense Glass Forming Brownian Suspensions After Flow Cessation}



\author{Anoop Mutneja}
\affiliation{Department of Materials Science and Engineering, University of Illinois, Urbana, IL 61801, USA}
\affiliation{Materials Research Laboratory, University of Illinois, Urbana, IL, 61801, USA}
\author{Kenneth S. Schweizer}
\email{kschweiz@illinois.edu}
\affiliation{Department of Materials Science and Engineering, University of Illinois, Urbana, IL 61801, USA}
\affiliation{Materials Research Laboratory, University of Illinois, Urbana, IL, 61801, USA}
\affiliation{Department of  Chemistry, University of Illinois, Urbana, IL 61801, USA}
\affiliation{Department of  Chemical \& Biomolecular Engineering, University of Illinois, Urbana, IL 61801, USA}


\date{\today}

\begin{abstract}
The re-solidification of amorphous solids after mechanically driven yielding from a nonequilibrium state is a fundamental soft matter science problem of broad relevance in materials science, with implications for material strength, processing, and printing-based additive manufacturing. We present a microscopic statistical mechanical theory that predicts in a unified manner the coupled time evolutions of structural and stress recovery following shear cessation from a mechanically prepared nonequilibrium state. The approach is built on recent advances in understanding activated dynamics in Brownian systems under both quiescent and startup continuous shear conditions. A particle-level microrheological model framework self-consistently incorporates stress generation, constraint softening due to external mechanical forces and structural deformation, and nonequilibrium memory effects. After flow cessation, the theory captures the re-building of kinetic constraints and activation barriers over time that underlie structural recovery, stress relaxation, and re-solidification through dynamic relaxation and an elementary form of convective elastic backflow. The ideas are general for particle-based materials, and quantitatively applied to dense hard-sphere Brownian colloidal suspensions which also serve as a foundational paradigm for glass forming materials where thermal fluctuations are important. The theory properly captures the rich range of stress relaxation behaviors observed experimentally that evolve from exponential, to stretched exponential, to fractional power law in form with increasing packing fraction. A microscopic understanding is achieved of the emergence of apparent residual stresses on practical laboratory timescales, power-law endless aging, sigmoidal recovery of the elastic modulus, pre-shear-rate-dependent memory effects, and a two-step structural relaxation process that can become decoupled from stress relaxation. 
\end{abstract}


\maketitle 

\section{Introduction}
\vspace{-0.3cm}
Kinetically arrested soft matter (glasses, gels) of diverse chemistry and material types, such as colloids and polymers, can be driven to undergo an amorphous solid-to-fluid transition when subjected to a sufficiently large external mechanical force. This process is broadly viewed as a nonequilibrium yielding transition, and is of both fundamental science interest and of high importance in materials processing and fabrication \cite{1noauthor2021theory-0e4,2bonn2017yield-35d,3barrat2011heterogeneities-180,4joshi2018yield-302,5berthier2025yielding-728,6nicolas2018deformation-6dc,7leishangthem2017yielding-897}. Upon removal of the external force or “flow cessation”, the system typically re-solidifies via nonequilibrium dynamics, which is critical to control for ultimate material properties \cite{8choi2020optimal-644}  and printing-based manufacturing \cite{9truby2016printing-8f7} . However, recovery dynamics is deformation history dependent with mechanical memory effects, and the re-equilibration can be exceptionally slow. The temporal structural, dynamical, and stress and strain evolutions are highly coupled. A microscopic predictive understanding of both the universal and materials-specific effects remains a major theoretical challenge, and is the topic of the present article.

Time-dependent physical aging following the removal of mechanical forces is influenced by how close the material is to its quiescent kinetically arrested state. In so-called endless “full aging” systems, the relaxation time grows \textit{linearly} with time after deformation cessation, with no equilibrated state attained on practical timescales \cite{2bonn2017yield-35d,4joshi2018yield-302,10masri2009dynamic-ab1,11bellour2003aging-96b}. However, for some systems such as colloidal fractal and depletion gels \cite{12cipelletti2000universal-f8b,13chung2006microscopic-344,14viasnoff2002how-957}, micellar polycrystals \cite{15cipelletti2002universal-c1e}, and colloidal and polymer glasses \cite{14viasnoff2002how-957,16li2019physical-67e,17mckenna2009soft-b45,18struik1978physical-6e8,19mckenna2003mechanical-7ec}, quiescent aging after a not too deep quench into a nonequilibrium state exhibits a sigmoidal temporal response corresponding to an intermediate time fractional power-law “sub-aging” growth of the relaxation time, followed by saturation signaling equilibration \cite{4joshi2018yield-302,14viasnoff2002how-957,15cipelletti2002universal-c1e,17mckenna2009soft-b45,18struik1978physical-6e8,19mckenna2003mechanical-7ec}. 

Nonequilibrium recovery phenomena are often monitored by measuring stress relaxation as a function of time after termination of the deformation. For fluid-like soft matter systems far from their quiescent glass transition, the stress decay is typically exponential, but becomes much slower, with a stretched exponential or even a fractional power law form, as the glass or other kinetic arrest transition is approached \cite{17mckenna2009soft-b45,20ballauff2013residual-748,21jacob2019rheological-986,22chen2020microscopic-96e,23pamvouxoglou2021stress-937}. The dynamics can become so slow that on experimental timescales re-equilibration is impossible, resulting in the observation of  apparent ``residual stresses" on practical timescales  \cite{17mckenna2009soft-b45,20ballauff2013residual-748,21jacob2019rheological-986,22chen2020microscopic-96e,23pamvouxoglou2021stress-937,24moghimi2017residual-90e}. In metastable Brownian suspensions (including the present study), the stress typically continues to decay, albeit extremely slowly, and thus the residual stress in our analysis denotes the unrelaxed stress within experimentally accessible times, and no true plateau out to infinite time is present. In contrast, for jammed non-Brownian suspensions \cite{25mohan2015buildup-d4d,26mohan2013microscopic-571,27vasisht2022residual-eef,28vinutha2024memory-113,29lidon2017powerlaw-024,30negi2010physical-17c}, such as emulsions, foams, and microgels, stress relaxation follows a two-step process: an initial rapid decay followed by a much slower relaxation that ultimately approaches a well-defined residual stress plateau. In some systems, this plateau can be followed by a subsequent stress increase \cite{31sudreau2022residual-fca,32murphy2020memory-675}, resulting in non‑monotonic stress relaxation, which has been attributed to shear‑localization phenomena, such as shear banding \cite{33ward2025shear-eb0}. The latter spatially inhomogeneous effect is not present in our approach which assumes spatially homogeneous deformation. 

Stress relaxation following shear cessation exhibits pronounced history dependence and memory effects. For example, the apparent residual stress depends on pre-shear rate in Brownian colloid glasses and non-Brownian jammed suspensions, a specific type of mechanical memory \cite{17mckenna2009soft-b45,20ballauff2013residual-748,21jacob2019rheological-986,22chen2020microscopic-96e,23pamvouxoglou2021stress-937,24moghimi2017residual-90e,25mohan2015buildup-d4d,26mohan2013microscopic-571,27vasisht2022residual-eef,28vinutha2024memory-113,29lidon2017powerlaw-024,30negi2010physical-17c}.  Moreover, stress relaxation profiles from materials yielded at different shear rates collapse reasonably well when the elapsed time is scaled by the corresponding shear rate, despite the system not being under active deformation \cite{20ballauff2013residual-748,23pamvouxoglou2021stress-937,24moghimi2017residual-90e,26mohan2013microscopic-571,27vasisht2022residual-eef,28vinutha2024memory-113}. This reflects a fascinating imprinting of the strain-rate dependence of the prepared non-equilibrium state on the post-cessation structural and/or stress-relaxation time. We note that in non-Brownian systems, deformation memory and subsequent aging have also been attributed to particle-level directional stress-distributions \cite{34edera2025mechanical-e3f,35bantawa2025stress-a91}.  Furthermore, if flow cessation is implemented at progressively increasing accumulated strains that span the range from the elastic, to anelastic, to the stress overshoot, and post-yielding regimes, the stress relaxation exhibits an incubation time which decreases with increasing deformation, perhaps linked to the structural and/or mechanical relaxation time of the corresponding non-equilibrium state \cite{21jacob2019rheological-986,22chen2020microscopic-96e}.

Beyond the above issues, how the nonequilibrium material structure evolves after flow cessation relative to the mechanical behavior is of high interest, though less studied experimentally and theoretically. With advancements in rheo-optics, it has been shown that, after shear cessation, soft glassy materials display convective‑like particle motion opposite to the direction of the prior shear, referred to as “convective backflow”, which can persist for long times \cite{22chen2020microscopic-96e}. In these experiments, particle displacements were measured following shear cessation under conditions of fixed macroscopic strain, while monitoring the progressively decreasing stress required to maintain that strain. The resulting displacements exhibited a strong component aligned with the direction \textit{opposite} to the prior applied shear. This effect can drive stress reduction \textit{elastically}, as confirmed experimentally \cite{22chen2020microscopic-96e}  by an observed correlation between backflow characteristic velocity and rate of stress decay. This additional relaxation exists in parallel with stress relaxation via thermally activated processes in Brownian suspensions. Another open question is whether there is a significant timescale difference between nonequilibrium structural and stress relaxations. Scattering measurements \cite{36denisov2013resolving-853}  on ultra-dense glassy repulsive colloids \textit{have} found that structural relaxation proceeds much more slowly than stress relaxation following shear cessation. 

There are no microscopic predictive theories formulated at the level of particles and forces for the coupled phenomena discussed above that include activated dynamics and their causal connection to structure and interactions. Diverse parametrized phenomenological models with variable degrees of predictive power exist, which address some of these phenomena, generally involving multiple fit parameters. For example, the “soft glass rheology” (SGR) model \cite{37sollich1997rheology-00d,38sollich1998rheological-2e5,39kaushal2016analyzing-494}, or its quiescent “trap model” analog \cite{40bouchaud1992weak-411,41monthus1999models-c48}, based on thermal and/or mechanical-noise-activated hopping out of a chosen quenched distribution of trap barriers, has been useful in bringing high level insights for soft matter systems \cite{4joshi2018yield-302,39kaushal2016analyzing-494,42moller2009attempt-34e,43fielding2008shear-69b}, albeit without any microscopic underpinning to interactions, structure, and thermodynamic state. The first microscopic approach that attempted to predict stress relaxation after shear cessation is Mode Coupling Theory (MCT), which is built on the idealized concept of a literal kinetically arrested solid with \textit{no} ergodicity-restoring activated processes \cite{20ballauff2013residual-748,44fritschi2014modecoupling-ecd}. It predicts shear-rate dependent residual stresses for hard sphere fluids above (below) the ideal MCT glass transition packing fraction (temperature), while the stress smoothly goes to zero in the putative fluid state. The obtained relaxation profiles \cite{20ballauff2013residual-748} qualitatively deviate in form from the experimental behavior mentioned above (and discussed below), and the shear rate dependence of the glassy regime residual stress is not captured \cite{20ballauff2013residual-748}. The latter was subsequently modeled by adding a phenomenological term to the empirically parameterized schematic version of ideal MCT\cite{44fritschi2014modecoupling-ecd}.

In this article we formulate a microscopic nonequilibrium statistical mechanical theory of coupled aging, activated structural relaxation, elasticity, and stress evolution following shear cessation for Brownian systems with thermal fluctuations. It builds on our recent efforts for the continuous step shear transient and steady state response of dense hard sphere \cite{45ghosh2023microscopic-39e} and attractive sphere suspensions \cite{46mutneja2025microscopic-f31}, including the issue of single versus double yielding. The approach formulated here is general for particle systems, including dense gels and attractive glasses with strong attractions that form physical bonds. Here we apply the ideas to the simplest glass-forming system -- ultra-dense hard sphere colloid suspensions. The new conceptual elements involve a self-consistent treatment within a microscopic generalized Maxwell model framework for stress generation, wavevector advected structural deformation, stress-induced kinetic constraint release, nonlinear elasticity, and activated structural and stress relaxation. After reaching a predicted non-equilibrium steady state, the external shear rate is set to zero, and the stress and structural relaxation profiles are predicted. The experimental protocol corresponding to our theoretical description involves a metastable Brownian suspension subjected to step‑rate shear, which is terminated at a time (subsequently re-defined as “zero” for post-cessation dynamics) corresponding to a prescribed accumulated strain that may lie in the elastic, overshoot, or steady‑state flow regime of the step-rate shear deformation. After cessation, the macroscopic strain is held fixed (zero imposed strain rate), and the stress required to maintain this condition is recorded as a function of time.

As a preview of our core new results, the theory properly predicts the experimentally observed \cite{17mckenna2009soft-b45,20ballauff2013residual-748,21jacob2019rheological-986,22chen2020microscopic-96e,23pamvouxoglou2021stress-937}  \textit{qualitatively} different stress relaxation profiles as a function of increasing particle concentration in the metastable dense regime, which range from simple exponential to slower stretched exponential with variable stretching exponents, and ultimately to extremely slow power-law behaviors with state-dependent fractional scaling exponents. Structural relaxation is predicted with and without a “convective backflow” process, which can accelerate stress relaxation and lead to two-step structural relaxation. Deformation-induced memory effects as encoded in the pre-shear rate dependence of the relaxation profiles and residual stress are analyzed, and the predicted trends (obtained with \textit{no} fitting parameters) are shown to be consistent with experiment. The full nonequilibrium aging or re-equilibration process after shear cessation is addressed, and near full aging behavior is predicted as a consequence of how the microscopic, short-length scale cage and larger length scale collective elastic barriers for structural relaxation slowly evolve towards equilibrium with time after deformation cessation.
\section{Background Theoretical Elements}
\begin{figure*}[t]
    \centering
    \includegraphics[width=0.850\textwidth]{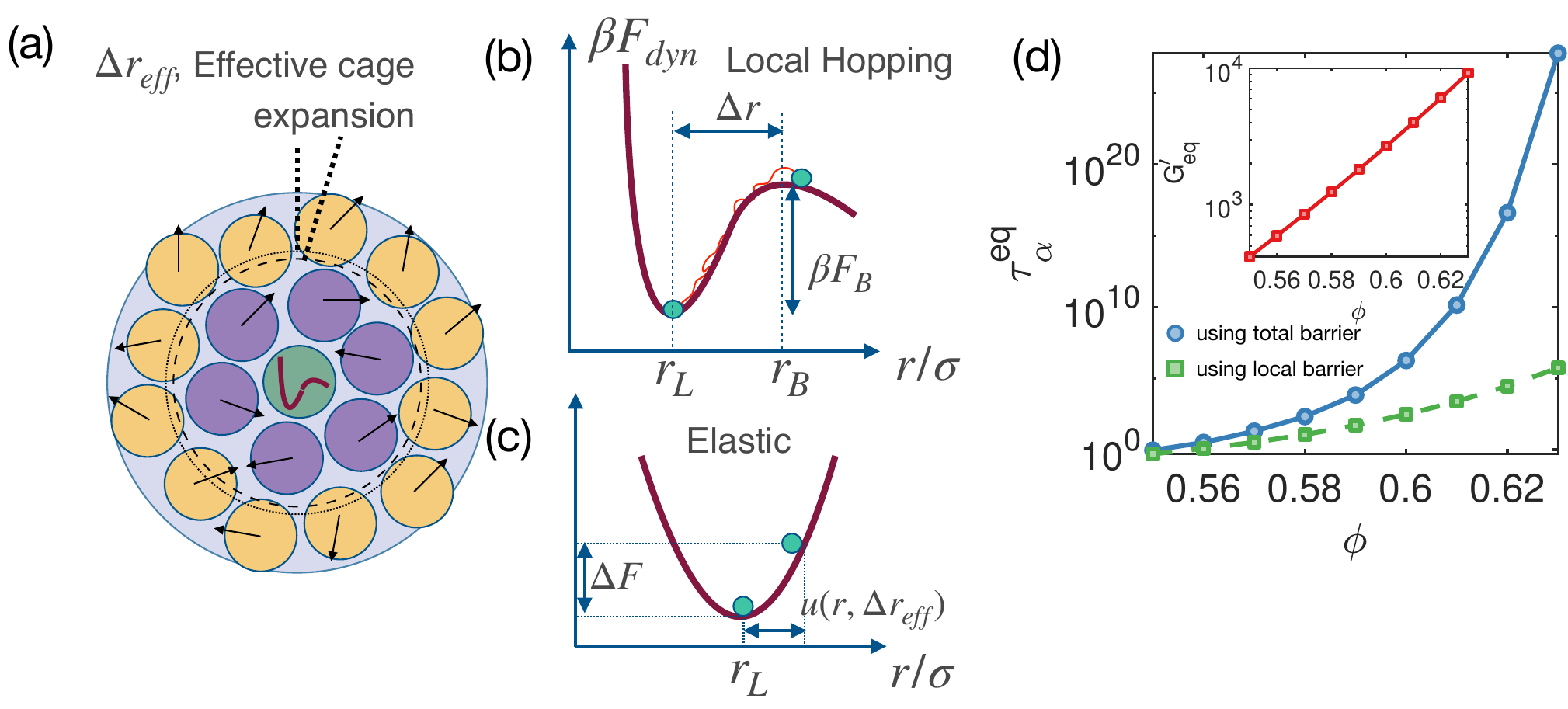}
    \caption{\textbf{Elements of the Quiescent Theories.} (a) The structural alpha relaxation event in ECNLE theory involves causally coupled cage scale large amplitude particle hopping and longer range small collective elastic displacements of all particles beyond the cage scale. (b) Dynamic free energy as a function of particle displacement in units of particle diameter ($\sigma$) with important length and energy scales indicated where $\beta=\left(k_BT\right)^{-1}$. (c) Small harmonic collective elastic displacement of all particles outside the cage with an amplitude $\mathrm{\Delta}r_{eff}$ nucleated at the cage boundary, and the associated elastic barrier at a distance $r$ from the cage center $\Delta F(r)$,  are indicated. (d) The dimensionless mean activated alpha relaxation time (and its analog with no elastic barrier, dashed curve) for dense metastable hard sphere fluids as a function of packing fraction. The inset shows the exponential growth of the dimensionless elastic shear modulus (units of $k_BT\sigma^{-3}$ ) with packing fraction associated with the transiently localized state, which has been recently experimentally verified \cite{57athanasiou2025probing-0e7,58chaki2024theoretical-92e} . }
    \label{fig1}
\end{figure*}

\vspace{-0.3cm}
For the benefit of the reader, we first recall key background foundational theoretical ideas concerning quiescent activated dynamics and the elementary aspects of their generalization to out-of-equilibrium systems under deformation. This has all been thoroughly documented in the literature, and predictions based on the theories have been extensively and successfully tested against experiments, including in our recent two JoR articles \cite{45ghosh2023microscopic-39e,46mutneja2025microscopic-f31}. We thus  present a high-level, largely physical description, with additional key technical details provided in the Appendices, and in the original references cited throughout the text. Those readers not interested in this background and the foundational theoretical elements discussed in Fig.\ref{fig1} can skip to Sec. \ref{Sec3} where the new theoretical development of this article for relaxation after shear cessation is formulated.

\subsection{Quiescent State Activated Dynamics}
\vspace{-0.3cm}
The Elastically Collective Nonlinear Langevin Equation (ECNLE) theory \cite{47mirigian2014elastically-4c6,48schweizer2005derivation-c99}  describes spherical single particle activated motion in dense fluids or suspensions at the stochastic trajectory level. The foundational idea is that slow density fluctuations are the physical origin of transient particle localization (“caging” for repulsive colloids), emergent intermediate time dynamic rigidity, and activated structural relaxation, and also control slow stress relaxation. This perspective allows one to derive via a stochastic evolution equation of a nonlinear Langevin equation (NLE) form \cite{48schweizer2005derivation-c99} (Appendix \ref{Apd:A1}) based on a novel form of non-ensemble-averaged dynamic density functional theory (DDFT) that encodes the idea of dynamic local equilibrium. The NLE microscopically relate interactions, structural pair correlations ($g(r)$ or its Fourier space analog $S(k)$, the structure factor) and thermodynamic state via an effective force exerted on each particle by the surrounding dynamically moving identical particles. This derived effective force can be represented as the negative gradient of a self-generated, \textit{spatially resolved}, dynamic free energy, $F_{dyn}\left(r\right)$, as a function of particle-displacement ($r$) as given in Eq.(\ref{eqn:1}) below. The dynamic free energy quantifies kinetic constraints on particle motion as a function of particle displacement [Fig.\ref{fig1}(a)] and is fully specified by $S(k)$ which is the origin of the predictive nature of the approach: 
\begin{equation}\label{eqn:1}
\begin{split}
    \beta F_{dyn}\left(r\right)=&-3\ln\left(\frac{r}{\sigma}\right)\\&-\frac{\rho}{2\pi^2}\int_{0}^{\infty}{\frac{k^2C^2\left(k\right)S\left(k\right)}{1+S^{-1}\left(k\right)}e^{-\frac{k^2r^2}{6}\left(1+\frac{1}{S\left(k\right)}\right)}}dk,
\end{split}
\end{equation}
where $\beta\equiv\left(k_BT\right)^{-1}$ is the thermal energy and $\rho$ is the fluid number density. The integration over wavevector reflects the summing up of the slowly decaying components of force-force time correlations on all length scales within a stochastic DDFT framework, in qualitative analogy with ideas adopted in microscopic mode coupling theory \cite{49gotze1992relaxation-8eb}.\\

For the specific hard sphere system studied here, the structure factor is obtained from the highly accurate Ornstein-Zernike (OZ) integral equation theory with the modified Verlet closure \cite{50zhou2020integral-c4b}  where the direct correlation function $C\left(k\right)=\rho^{-1}\left[1-S^{-1}\left(k\right)\right]$, $\phi\equiv\rho\pi\sigma^3/6$ is the packing fraction, and $\sigma$ the particle diameter. An example of the strongly anharmonic dynamic free energy in the deeply metastable regime is shown in Fig. \ref{fig1}(b) and displays the two characteristic features of strong caging. The first is a minimum at a small displacement $r_L$, indicating transient particle localization on a length scale much less than particle size. This feature controls dynamic solidity and the high-frequency shear modulus, $G^\prime$, as computed using a standard Green-Kubo-like microscopic relation that relates stress and density correlations \cite{51nagele1998linear-c57}  (see Appendix \ref{Apd:A2}, Eq.(\ref{eqn:GPrime})). The second key feature is the local cage barrier $F_B$ at a particle displacement $r_B$ that is critical for predicting thermally activated viscous dynamics (see Fig.\ref{fig1}(b)). For hard spheres, this cage barrier is finite for all packing fractions below random close packing (RCP). However, the predicted jump distance required to escape the cage, $\Delta r\equiv r_B-r_{loc}$, is sufficiently large that a small amount of extra local space is physically required, achieved via a coordinated spontaneous dynamical cage expansion. The inclusion of this nonlocal part of the alpha relaxation process defines ECNLE theory and has been shown to be of essential importance for understanding deeply metastable suspensions and viscous glass forming liquids \cite{47mirigian2014elastically-4c6}.  Conceptually, it is causally associated with the cage scale hopping event, and introduces a spatially longer range collective elastic barrier $F_{el}$ associated with the correlated elastic displacement field of particles outside the cage [Figs.\ref{fig1}(a) and \ref{fig1}(c)] which is \textit{fully} quantified from the same dynamic free energy \cite{52dyre2006elastic-c3c} (see Appendix \ref{Apd:A3}).

The average structural relaxation time is then predicted using Kramers mean first passage time for barrier crossing in a high friction viscous environment \cite{53kramers1940brownian-150} as a surrogate for the structural or alpha relaxation time, $\tau_\alpha$ (Eq. (\ref{eqn:TauAlpha})). A distribution of barriers and relaxation times based on local fluctuations of structure can be determined \cite{54schweizer2004activated-f43,55xie2020collective-e13}, but in the present minimalist microscopic framework is not considered. Physically, under strong deformation it is known that quiescent dynamic heterogeneity effects are strongly suppressed; see, for example, Ref.\cite{56hebert2015effect-983}.

This completes the overview of the conceptual elements of the quiescent equilibrium ECNLE theory. In all our results and discussions below, the elastic modulus and stress are in Brownian units of $\frac{k_BT}{\sigma^3}$, and the relaxation time and inverse applied strain rate are in units of the in-cage non-activated local dynamic process time scale $\tau_s$ (see Appendix:\ref{Apd:A1}, Eq(\ref{eqn:Taus})).

Fig.\ref{fig1}(d) shows mean alpha time and elastic modulus calculations as a function of packing fraction in the dense metastable regime of hard spheres. The cage (elastic) barrier in thermal energy units grows enormously from $4.7$ to $20.1$ ($0.6$ to $50.3$) for $\phi=0.55$ to $0.63$.  Hence, the elementary quiescent relaxation event becomes increasingly more spatially nonlocal and collective with densification. Importantly, the predicted exponential growth of the shear modulus with packing fraction is in excellent agreement with recent colloid experiments \cite{57athanasiou2025probing-0e7,58chaki2024theoretical-92e}.
\subsection{Mechanically prepared Nonequilibrium State} 
\vspace{-0.3cm}
To study the problem of nonequilibrium dynamics after shear cessation from a steady state based on a constant step-rate ($\dot{\gamma}$) shear deformation first requires a theoretical description of the prepared non-equilibrium steady state. The extension of ECNLE theory to address this problem has been thoroughly discussed in prior articles \cite{45ghosh2023microscopic-39e,46mutneja2025microscopic-f31}, and its quantitative predictions have been successfully confronted with experiments on dense hard sphere metastable colloidal suspensions. Here, we briefly review this foundational work for the benefit of the reader. We first provide a physical description of the theoretical ideas, and then summarize their precise mathematical formulation. The central idea is to relate the macroscopic rheological variables and response to the microscopic particle trajectory, forces, and structure under active deformation. It is in this sense the approach is akin to a microrheology perspective, and is the origin of its predictive power.

The rheological response to step-rate ($\dot{\gamma}$) shear deformation is quantified by the acquired macroscopic stress, $\Sigma\left(\tilde{t}\right)$, and an effective \textit{microscopically-defined} strain variable, $\gamma_{eff}(\tilde{t})$, where $\tilde{t}$ is the elapsed time after the imposition of shear and $\Sigma=\gamma_{eff}=0$ in the quiescent state. The physical meaning of the acquired stress is self‑evident, whereas $\gamma_{eff}\left(\tilde{t}\right)$ encodes a key novel physical aspect of our theoretical approach. Specifically, as precisely defined below, it is the recoverable portion of the total strain, $\gamma$, stored in the nonequilibrium microstructure via the deformation‑induced modification of the structure factor \cite{59kamani2025linking-e10}. This effective strain equals the macroscopic mechanical total strain for small $\gamma$ per an affine elastic response, but then deviates and ultimately saturates in the steady state due to viscous or plastic processes consistent with the attainment of a steady-state deformed microstructure. Crucially, stress acquisition and structural deformation \textit{both} induce weakening of microscopic kinetic constraints as encoded in the nonequilibrium analog of the dynamic free energy, leading to reductions in the elastic shear modulus and relaxation time, but for physically distinct reasons. Internally consistent with the single particle basis of the activated ECNLE theory of activated dynamics, the external stress is conceptually taken to be transduced down to the particle‑level as an external driving force that directly enters the nonequilibrium anharmonic dynamic free energy as an extra term of the form of a mechanical work. This stress-induced force increases the particle localization length thereby weakening the elastic shear modulus, and lowers the activation barrier thereby accelerating the activated relaxation process. In contrast, structural deformation as quantified by $\gamma_{eff}$, which for hard spheres can be qualitatively thought of as cage stretching, alters the kinetic constraints encoded in the dynamic free energy on all length scales (wavevectors, $k$) via the structure factor in the integrand of the second interparticle force based term in Eq(\ref{eqn:1}). This structural deformation also reduces of all localizing features of the anharmonic dynamic free energy, reducing the  elastic shear modulus and activation barrier, albeit in a manner distinct from the stress-induced force on a particle. Accordingly, the ECNLE framework is generalized such that the fundamental dynamic free energy, and hence $G^\prime$ and $\tau_\alpha$, become \textit{functionals} of the instantaneous value of $\Sigma$ and $\gamma_{eff}$ in a rheological experiment.

We now sketch the technical statistical mechanical realization of the above ideas.  Macroscopic stress enters as an effective force on each moving particle in a microrheological spirit, or equivalently per the NLE equation-of-motion as an additional microscopic mechanical work term in the now nonequilibrium dynamic free energy\cite{60kobelev2005strain-2be}: $\beta F_{dyn}\left(r,{\Sigma}\right)=\beta F_{dyn}\left(r,\mathrm{\Sigma}=0\right)-\frac{\pi\sigma^2}{24}\Sigma r$ . See Appendix \ref{Apd:B} for a more detailed discussion, including pedagogically valuable quantitative consequences in the \textit{hypothetical} case where the equilibrium structure is held fixed and stress is applied, and or only structure is deformed but there is no stress-induced force on a particle (Fig.\ref{fig12}). The deformed structure is microscopically modelled via the wave-vector advection idea of Fuchs and Cates \cite{61fuchs2002theory-373}  as proposed in their ideal MCT-ITT (integration through transients) analysis of hard sphere fluid rheology within the computationally tractable simplified “isotropically sheared” framework \cite{62amann2014transient-b79,63fuchs2009high-399} : $S\left(k,\gamma_{eff}(\tilde{t})\right)\rightarrow S\left(k\left[1+\left(\frac{\gamma_{eff}^2\left(\tilde{t}\right)\ }{3}\right)\right]\right)$. Crucially, this description connects the time evolving effective mechanical strain to microscopic structure and dynamics by “advection” of the structure factor to smaller wavevectors (which reduces kinetic constraints) while preserving its functional form. In our work, this idea involves the introduction of \textit{no} adjustable parameters, and hence microscopic predictability is preserved. Moreover, as a result, the effective strain can be expressed in terms of the time‑dependent advected wavevectors $k(\tilde{t})$ and their equilibrium values $k$: $\gamma_{eff}\left(\tilde{t}\right)\equiv\sqrt3\left[\left(\frac{k}{k\left(\tilde{t}\right)}\right)^2-1\right]^{1/2}$. Since all wavevectors evolve concurrently in order to preserve the shape of $S\left(k\right)$, for simplicity reasons only we use in our discussions and plots the location of the first peak of $S\left(k\right)$ at ${k=k}^\ast$ (commonly called the “cage order parameter”) as a representative scalar measure of the structural evolution. In real space, the near contact magnitude of the radial distribution function $g(r)$ (see Fig.\ref{fig13}(a)-inset) is reduced consistent with the physical idea that wavevector advection “softens” structural packing and cage correlations. This effect in turn reduces the activation barriers (both cage and elastic) and relaxation time, and increases the localization length thereby reducing the elastic shear modulus. We emphasize that the nonequilibrium dynamic free energy germane to rheology is now a function of both stress and effective structural strain.

The adoption of an averaged isotropic description of structural deformation has been done largely for practical reasons since accurate theoretical knowledge of the anisotropic structure factor based on integral equation theory is largely nonexistent. Moreover, the inclusion of a vectorial dependence of structure would render the construction of kinetic constraints in the dynamic free energy far more complex both conceptually and computationally. We do note that this simplification has simulation \cite{64koumakis2012yielding-592}  and experimental \cite{65besseling2007threedimensional-9cf}  support on the primarily local length scales relevant in the dynamical theory that enter via the dynamic free energy. From a tensorial perspective, the associated elastic and viscous softening effects due to structural deformation are attributed primarily to extensional-axis-dominated distortions, as expected based on simulation of the local anisotropy of $g(r)$ under deformation \cite{64koumakis2012yielding-592,66marenne2017unsteady-1ac}. Of course, the isotropic approximation may break down at very high strain rates, where rapid densification along the compression axis can generate non‑trivial effects \cite{67pan2023review-d54,68koumakis2013complex-b0e}(e.g., they underlie so-called shear jamming), but addressing this is an unsolved open question in the present microscopic theoretical approach.

 The above theoretical approach allows prediction of the elastic shear modulus and activated relaxation time for any prescribed combination of applied or internally generated stress and structural deformation. The remaining task is to determine the nonequilibrium stress–strain response under imposed active deformation. As discussed in great detail and successfully employed to understand both dense colloidal \cite{45ghosh2023microscopic-39e,46mutneja2025microscopic-f31}  and polymer glass rheology \cite{69chen2011theory-f68}, we adopt the simplest suitable framework we can envision: a generalized Maxwell model characterized by a single \textit{nonequilibrium} elastic modulus and relaxation time that evolves with deformation:

\begin{equation}\label{eqn:2}
    \frac{d\Sigma\left(\tilde{t}\right)}{d\tilde{t}}=-\frac{\Sigma\left(\tilde{t}\right)}{\tau_\alpha\left[\Sigma\left(\tilde{t}\right),\gamma_{eff}\left(\tilde{t}\right)\right]}+\ \dot{\gamma}{\left(\tilde{t}\right)G}^\prime\left[\Sigma\left(\tilde{t}\right),\gamma_{eff}\left(\tilde{t}\right)\right],
\end{equation}
\begin{equation}\label{eqn:3}
    \frac{dk\left(\tilde{t}\right)}{d\tilde{t}}=-f\frac{k\left(\tilde{t}\right)-k}{\tau_\alpha\left(\mathrm{\Sigma},\gamma_{eff}\right)}+\left(\frac{-\dot{\gamma}k\gamma}{3\left(\gamma^2/3+1\right)^{1.5}}\right)_{\gamma=\gamma_{eff\left(\tilde{t}\right)}}.
\end{equation}
Such a generalized Maxwell model is the simplest possible description since it includes only the elastic modulus and the mean relaxation time associated with slow activated dynamics out of the localized cage state. The word “generalized” Maxwell refers to the present theory microscopically accounting for the nonequilibrium time-dependent evolution of both the elastic modulus and structural relaxation time with time, stress, shear rate, and strain, treated in a \textit{self-consistent} manner. This is the origin of the richness of the theoretical prediction made here and in our prior work \cite{45ghosh2023microscopic-39e,46mutneja2025microscopic-f31}.

The physical content of Eqs (\ref{eqn:2}) and (\ref{eqn:3}) is as follows. The system acquires stress at the time varying elastic rate $\dot{\gamma}G^\prime(\tilde{t})$ and dissipates it exponentially on the timescale $\tau_\alpha\left(\tilde{t}\right)$ that also evolves with elapsed time. Likewise, the wavevectors that enter the deformed structural evolution follow the advection process at the affine rate $\dot{\gamma}$ to deform the structure at a strain $\gamma_{eff}$. The last term in Eq.(\ref{eqn:3}) corresponds to the differential form of the driving wavevector advection process where structure is deformed via the effective strain $\gamma_{eff}\left(\tilde{t}\right)$, while simultaneously attempting to relax towards the equilibrium structure with an  active deformation (and hence time evolving) dependent timescale $\frac{\tau_\alpha(t)}{f} $. The relaxation process is described as proceeding in the opposite direction to deformation, in a manner that the wavevectors that enter the structure factor are advected back to equilibrium. This is our theoretical description of what is referred to in experiments \cite{22chen2020microscopic-96e}  as convective backflow. It should  be contrasted with a classic activated relaxation process that does not invoke the shear-based advection idea. The relaxation terms in both equations capture activated plastic effects, whereas the growth terms describe elastic response.

As a secondary aspect of Eq.(\ref{eqn:3}), a constant numerical prefactor $f$ has been introduced to model the possible material-specific quantitative differences between the stress and structural relaxation times which are well known to exist in quiescent glassy materials \cite{36denisov2013resolving-853}. Although there is a clear physical reason for the presence of this factor, at present it is not \textit{a priori} predicted. We adopt the simplest $f=1$ choice in this work. We do note that experimental observations of quiescent glassy liquid relaxation \cite{36denisov2013resolving-853}  typically find $f<1$, i.e. structural relaxation is a modestly slower than stress relaxation probed in linear viscoelastic measurements, although the time scales follow essentially identical variations with temperature and density.  More generally, we believe this inequality can be physically intuitive if structural relaxation involves even modestly larger length scales than stress relaxation. This point is of highly second-order importance for our present work. It is briefly explored in the SI Fig.\ref{fig.S2}, and more generally reflects physics that remains debated even under quiescent conditions.

To summarize, Eqs. (\ref{eqn:2}) and (\ref{eqn:3}), plus the standard tools of ECNLE theory that allow the microscopic statistical mechanical computation of the elastic shear modulus and alpha relaxation time from the spatially-resolved dynamic free energy, yield a close mathematical description for the nonlinear rheology in the stress versus macroscopic accumulated strain ($\gamma$) representation for a given equilibrium structure factor, thermodynamic state, and shear rate. The approach is a highly nonlinear, self-consistent description that includes the full integration through transients of the nonequilibrium dynamics. The theory has previously been shown to successfully capture key experimental phenomena, such as the static yield stress overshoot, shear thinning, and steady-state flow curves in dense hard sphere colloidal suspensions as functions of packing fraction and shear rate \cite{45ghosh2023microscopic-39e}. For the present new work, the analysis for a step-rate continuous start-up stress-strain curve at a constant strain rate serves as the “preparation” step of the functionally coupled nonequilibrium structural and mechanical state.

\section{Structural Aging, Solidity Recovery, and Stress Relaxation}\label{Sec3}
\vspace{-0.3cm}
We now formulate a theory for structural and elastic solidity recovery and stress relaxation upon deformation cessation from a mechanically prepared nonequilibrium state. We believe that our use of the word “recovery” is an accurate descriptor, but it can also be viewed as “re-equilibration” to the initial quiescent state; it should not be confused with the use of the word recovery in creep experiments. The new ideas we develop below are general in that they apply if deformation is terminated in the nonequilibrium steady flowing state, or from any other region of the stress-strain curve. We focus almost exclusively on the first protocol, both for simplicity, and because most experiments and simulation studies follow this protocol. Some results are given for the second protocol in Sec. \ref{Sec7}. For post shear cessation which commences at (by definition) $t=0$, the elapsed time is indicated as $t$.

For structural evolution in a continuous step-rate shear deformation, the driving term appears via the last wavevector advection term in Eq.(\ref{eqn:3}), which will be absent for shear cessation experiments, and thus only the structural relaxation term is present: $\frac{dk\left(t\right)}{dt}=-f\frac{k\left(t\right)-k}{\tau_\alpha\left(t\right)}$. Given that the nonequilibrium structural relaxation evolves with time opposite to deformation, a net nonzero \textit{internal} strain rate is generated at intermediate times per $\frac{d\gamma_{eff}(t)}{dt}\equiv-\sqrt{\frac{3}{\left(\frac{k}{k\left(t\right)}\right)^2-1}}\frac{k^2}{k\left(t\right)^3}\frac{dk\left(t\right)}{dt}$. This quantity enters Eq.(\ref{eqn:2}) as a $microscopic$ proxy for $\dot{\gamma}(t)$. Crucially, this internal strain rate is \textit{not} akin to an external driving force via an applied macroscopic shear deformation. Rather, it is a physical dynamical variable that is directly related to the evolution of the nonequilibrium structure, and thus can be thought of as an internal recovery strain rate which elastically releases the stress in Eq.(\ref{eqn:2}). To provide further physical interpretation of this mechanism, consider the nonequilibrium steady state at time $t=0^-$ (just before shear cessation), where internal structural relaxation must balance the imposed deformation rate $\dot{\gamma}$ to sustain steady-state structure and flow. Immediately following shear cessation at $t=0^+$ (after cessation), this balance is disrupted, and the system experiences an unbalanced internal strain-rate of ${\dot{\gamma}}_{eff}\left(0^+\right)=-\dot{\gamma}$. The subsequent evolution of ${\dot{\gamma}}_{eff}\left(t\right)$ is governed by the so-modified Eq.(\ref{eqn:3}), and the resulting internally generated strain rate enters Eq.(\ref{eqn:2}) as an elastic contribution that reduces the stress. This provides our microscopic description of the experimentally \cite{22chen2020microscopic-96e}  observed elastic convective backflow process.

The coupled stress and structural deformation time-dependent recovery profiles then follow by self-consistently solving the so-modified Eqs.(\ref{eqn:2}) and (\ref{eqn:3}). The initial conditions are $\Sigma\left(t=0\right)=\Sigma\left(\tilde{t}\rightarrow\infty\right)\equiv\Sigma^{Int}$ and $k(t=0)=k(\tilde{t}\rightarrow\infty)$, which leads to $\gamma_{eff}\left(t=0\right)=\sqrt3\left[\left(k/k\left(t=0\right)\right)^2-1\right]^{1/2}\equiv\gamma_{eff}^{Int}$. The initial stress relaxation time $\tau_\alpha\left(0\right)=\tau_\alpha\left(\Sigma^{Int},\gamma_{eff}^{Int}\right)\equiv{\tau_\alpha}^{Int}$ and the elastic modulus $G^\prime\left(0\right)=G^\prime\left(\Sigma^{Int},\gamma_{eff}^{Int}\right)\equiv G\prime^{Int}$ then follow. 

The above formulation defines what we call the Model-A version of the new theory. Structural relaxation proceeds in an opposite manner to applied deformation within a simple description of convective elastic backflow \cite{22chen2020microscopic-96e}. Importantly, this latter process results in a partial \textit{decoupling} between stress and structural relaxation via an internally generated backflow that elastically reduces the stress.

We have also explored an alternative Model-B in which the convective backflow effect is neglected by setting $\dot{\gamma}\left(t\right)=0$ in Eq.(\ref{eqn:2}), and thus stress and structural relaxation become \textit{slaved}. Predictions based on Model-B are presented primarily to establish  the role of convective backflow within the context of the theory. We believe that Model‑A more faithfully captures the correct underlying physics. Any differences between the theoretical results can potentially be tested against experiments or simulations. Very importantly, we will demonstrate the stress relaxation predictions do \textit{not} qualitatively change depending on which model is adopted, \textit{but} qualitative differences are predicted for structural evolution.

Our theoretical framework incorporates aging (or re-equilibration back to the quiescent state) and activation barrier rebuilding after shear cessation. In \textit{qualitative} contrast to phenomenological models, these physical phenomena are treated microscopically, and key effects are not imposed manually, but rather emerge self-consistently. Recall that ideal MCT \cite{20ballauff2013residual-748,44fritschi2014modecoupling-ecd,62amann2014transient-b79,63fuchs2009high-399}  describes the consequences of flow cessation as a stress reduction via \textit{equilibrium} structure decorrelation dynamics, and does not capture deformation-induced shear melting, aging after cessation, and activated relaxation. These differences are crucial to the new physical behaviors predicted below.  
\begin{figure*}
    \centering
    \includegraphics[width=0.850\textwidth]{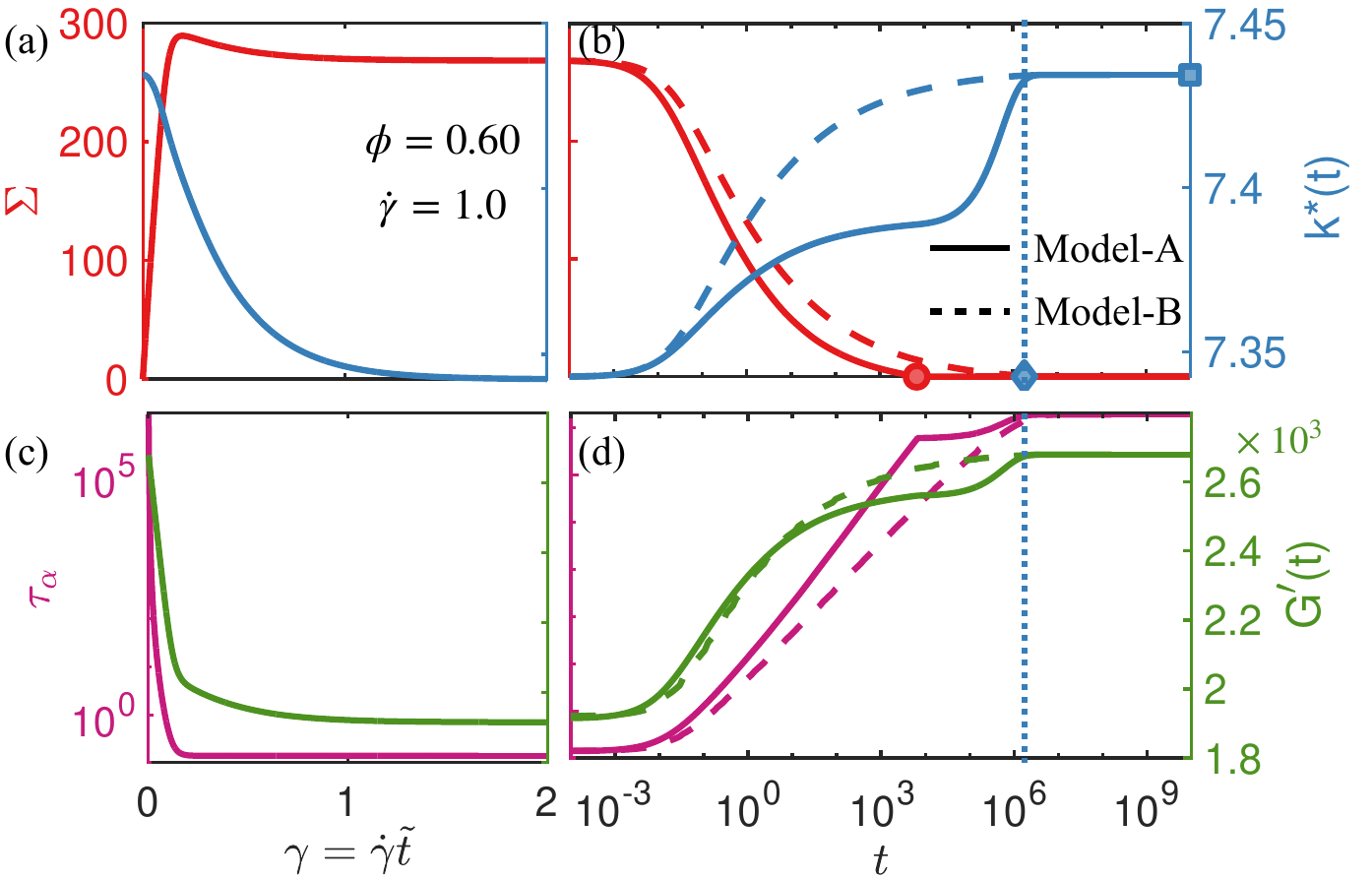}
    \caption{\textbf{Property Evolutions Under Continuous Start-up Shear and After Shear Cessation.} (a) Stress and structural order parameter (first peak location of $S(k=k^\ast)$ ) evolution during continuous start-up shear for $\phi=0.60$ at a dimensionless shear rate of $\dot{\gamma}=1.0$. (b) Recovery of stress and the structural order parameter post deformation cessation. Solid curves are results based on Model-A, dashed curves correspond to Model-B. (c) Evolution of the mean alpha time and elastic shear modulus during deformation. The dimensionless alpha time decreases to $0.14$ (with a nonzero but small barrier) in the flowing nonequilibrium steady state from the very large quiescent value of $\sim{10}^6$. (d) Rebuilding of the alpha time and elastic modulus post-cessation. The dotted curve and the blue diamond indicates the quiescent alpha time, and the red circle the time when stress reaches zero in Model-A.   }
    \label{fig2}
\end{figure*}
For the ultra-dense Brownian suspensions of present interest where activated dynamics is paramount, as in prior studies 45,46  the hydrodynamic interactions (HI) are modeled as only entering via the short elementary timescale, $\tau_s$, (Appendix.\ref{Apd:A1}) which is taken to be a deformation and aging independent non-dimensionalization timescale. It is associated with “in cage” dissipative short time and distance dynamics, successfully modeled previously using a non-activated binary collision mean field theory \cite{70cohen1998viscosity-7d3}. Alternatively, a beyond single particle local hydrodynamics model could be employed, but this will have very little effect on any of our results since this physics only enters as the time prefactor for activated dynamics. For a typical micron size colloid, $\tau_s$ is often of order seconds, but varies with temperature, solvent viscosity, and colloid radius in a well known manner. Effects such as longer range coherent HI-driven backflow \cite{71nakayama2005simulation-ab9,72mohanty2020transient-6b7} are absent, and should not to be confused with the presented convective elastic backflow description. Ultimately, our results can be tested against Brownian dynamic simulations and compared to those that include HI. Our approach is, in principle, applicable to all spherical particle systems with  any microscopic  interaction potential. 

\section{Evolution of Coupled Structure, Relaxation, and Elasticity for a Baseline System}
\vspace{-0.3cm}
We begin by presenting results for a representative “baseline” system defined as an initially prepared nonequilibrium flowing steady state with a fixed dimensionless shear rate of $\dot{\gamma}=1.0$ and a high $\phi=0.60$, with $f=1$ in Eq.(\ref{eqn:3}).  Equations Eq.(\ref{eqn:2}) and Eq.(\ref{eqn:3}) are solved for a step-rate start-up deformation to obtain the nonequilibrium flow state, after which the deformation is stopped and relaxation dynamics is studied using Model-A and Model-B. The focus is initially on four inter-related structural, dynamical, elastic, and mechanical properties (Fig.\ref{fig2}). 

Figures \ref{fig2}(a) show the time evolution of stress and structure during the step-rate start-up shear. Figures \ref{fig2}(c) illustrate the corresponding shear melting behaviour driven by the deformation-induced reduction of the structural alpha relaxation time and elastic modulus. We remind the reader that all wavevectors that enter via the structure factor evolve concurrently based on the adopted advection idea. Thus, as a simple and intuitive cage order parameter, one can describe structural evolution by the location of first peak of $S\left(k\right)$ at ${k=k}^\ast$. The x-axis of the left panel is the bare accumulated strain controlled experimentally, $\gamma=\dot{\gamma}\tilde{t}$. Panels (b) and (d) show the corresponding post-cessation stress and structure relaxation profiles, as well as the alpha time and modulus aging behavior. As an example, if the elementary short in-cage relaxation time is $\tau_s\sim3$ secs (a typical value for a micron-sized colloid at room temperature), the overall time range in the figure extends to ${10}^{10}\tau_s\sim100$ years, while the system relaxes in around ${10}^6\tau_s\sim36$ days. Hence, the results are relevant to the emergence of \textit{apparent} residual stress and nonequilibrium structure on practical experimental timescales. Of course, since ECNLE theory predicts the structural relaxation time is always finite for hard spheres below RCP, a residual stress as defined as a literally time-independent stress plateau out to infinite observation time does not exist.

Figure \ref{fig2} (b) contains the stress and structural relaxation predictions for both Model-A and Model-B. In Model-A, the structural relaxation generated internal strain-rate (via elastic convective backflow) contributes to stress reduction, and thus the structure and stress relaxations are decoupled to some extent. In contrast, in Model-B the stress and structural relaxations are driven solely by the time-evolving nonequilibrium relaxation time. As a result, the stress and structure relaxation profiles in Model-B are similar, reaching quiescent values at roughly the quiescent $\tau_\alpha$ indicated by the dotted vertical line. 

Apparent power law growth of the alpha time and sigmodal evolution of $G^\prime$ with elapsed time are predicted [Fig.\ref{fig2} (d)] for both Models A and B. These can be viewed as a type of physical aging in the sense of re-equilibration to the quiescent state after removal of external forces from a mechanically prepared nonequilibrium state. This is not the same “aging” process as occurs upon a rapid quench of a quiescent glass forming material via lowering temperature or increasing density \cite{17mckenna2009soft-b45,73peng2014comparison-631}, but rather denotes aging towards equilibrium from a mechanically prepared nonequilibrium state. Deformation in the pre-cessation steady state modestly reduces the elastic modulus, and massively reduces the alpha time thereby facilitating flow. Upon flow cessation, such aging begins as stress decreases and the structural recovery towards the quiescent state begins. This aging behaviour, in turn, self-consistently influences the stress and structure relaxation profiles. 

In Model-A, faster aging is predicted because of the presence of the additional stress relaxation mechanism. It leads to an interesting 2-step decay of the structural order parameter, and an apparent time-persistent ``frozen" structural deformation which only relaxes on the very long quiescent alpha timescale. This represents a memory effect of shear imprinted on the structure during the deformation preparation stage. Slower structural evolution compared to stress relaxation has been experimentally observed in dense repulsive colloidal suspensions (Denisov \textit{et al.} \cite{36denisov2013resolving-853} and McKenna \textit{et al.}\cite{30negi2010physical-17c}), although a microscopic explanation is absent. Our results provide a possible mechanistic theoretical basis. The amplitude of this frozen-in component of structure is very small in the dense colloidal suspension experiments \cite{36denisov2013resolving-853}, as also true in our results as encoded in $k^\ast\left(t\right)$. This behaviour stands in contrast to the larger effect on the elastic modulus, and the massive effect on the alpha time. An interesting prediction is the quiescent elastic properties are recovered only at very long times, essentially the quiescent alpha time which is often inaccessible for experimental glassy systems, \textit{even after} the internal stress has decayed to zero. 

In contrast to the structural recovery dynamics discussed above, the predicted stress relaxation profiles based on model-A and model-B are similar. In Secs. \ref{Sec5}-\ref{Sec7}, we present many additional results using model-A and a few based on model-B; additional results from model-B are given in the SI and are qualitatively similar except for a few noted differences. 

To summarize this section, Fig. \ref{fig2} illustrates our core messages for a single baseline choice of pre-shear rate and packing fraction, which we show below are qualitatively general. At a detailed level that is highly relevant to experiments and simulations, there are significant dependencies of the predicted quantities on these two controllable variables. The role of a possible mismatch in the evolution equations for stress and structural relaxation per $f\neq1$ in Eq.(\ref{eqn:3}) is discussed in Fig. S2 in the SI where it is argued this a second order issue.
\section{Role of Variable Degree of Quiescent Glassiness: Packing Fraction }\label{Sec5}
\vspace{-0.3cm}
The quiescent activated relaxation time is enormously sensitive to packing fraction in the metastable regime [Fig.\ref{fig1}(c)]. Moreover, the relative importance under quiescent conditions of the longer range collective elastic barrier (Appendix:\ref{Apd:A3}) is well known to increase greatly with packing fraction based on ECNLE theory \cite{47mirigian2014elastically-4c6}. How these changes impact the step-rate shear continuous rheology has been thoroughly treated previously \cite{45ghosh2023microscopic-39e,46mutneja2025microscopic-f31}, and the theory accurately captures all important experimental features. The new question is its role on nonequilibrium recovery kinetics of the baseline system.
\subsection{Re-equilibration dynamics}
\vspace{-0.3cm}
\begin{figure*}
    \centering
    \includegraphics[width=0.950\textwidth]{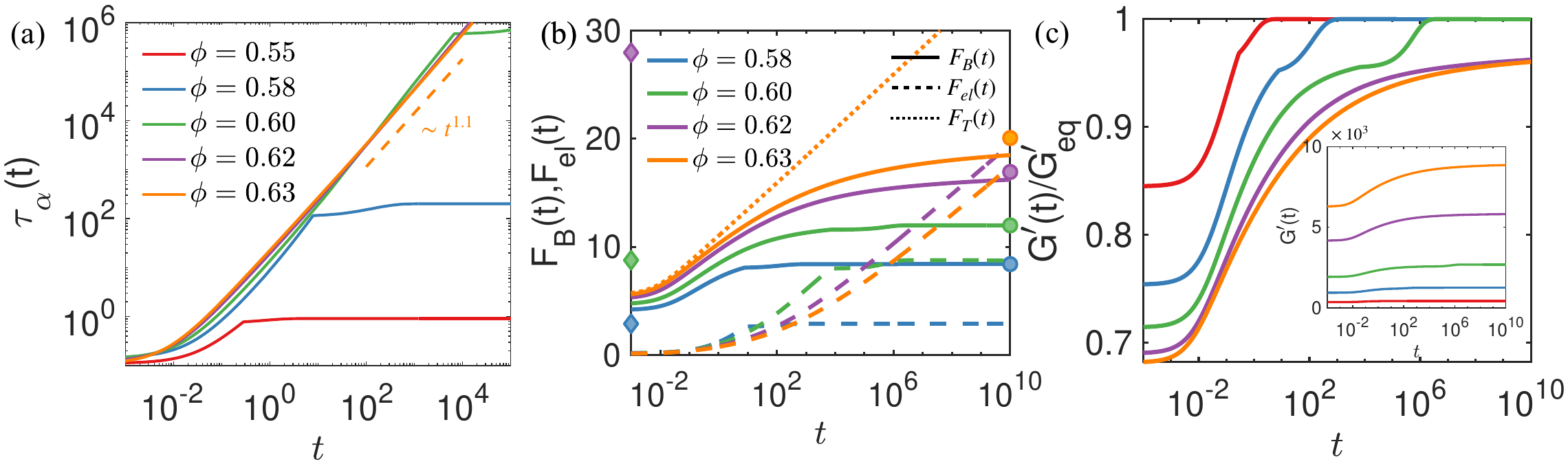}
    \caption{\textbf{Re-equilibriation Dynamics after Flow Cessation} (a) Rebuilding of the alpha time after flow cessation from the steady state of a continuous start-up shear deformation at $\dot{\gamma}=1.0$ for different packing fractions. (b) Rebuilding of the local (solid), elastic (dashed), and total (dotted) barriers in thermal energy units for different packing fractions corresponding to quiescent alpha times that vary by $\sim25$ decades as shown in Fig.\ref{fig1}(c). Circles on the right y-axis indicate the quiescent local barriers, while diamonds on the left y-axis represent the quiescent elastic barriers. (c) Elastic modulus rebuilding, scaled by the quiescent modulus in the main panel, with unscaled data shown in the inset.}
    \label{fig3}
\end{figure*}
We first consider the effects of post-shear cessation re-equilibration dynamics on the fundamental theoretical quantities as packing fraction is changed compared to the baseline value of $\phi=0.6$. Figure \ref{fig3}(a) shows the temporal growth of the alpha time for $\phi\in[0.55-0.63]$ after shear cessation. Initially, it is relatively small and pre-shear rate dependent, as will be discussed further below. But it then increases gradually, and becomes of an effective power-law form after an initial onset time, in a manner that depends on the alpha time of the prepared non-equilibrium steady state. Importantly, such \textit{apparent} power laws here and below do not have a simple “dynamic scaling” explanation. Rather, they are predictions that reflect the intricate, self-consistent, nonlinear evolution with time of the coupled elastic modulus, relaxation time, and rheological variables out of equilibrium that define the core elements of our theory. 

For lower packing fractions in the above range, the power law in Fig.\ref{fig3}(a) eventually saturates to the equilibrium alpha time. But for higher packing fractions, the power-law aging behaviour extends to extremely long times beyond that shown in the figure which in experiments or simulations would be unmeasureably long. The apparent power-law exponent evolves with the elapsed time after shear cessation (Fig.\ref{fig.S3}). It starts from zero at very early times, rapidly exceeds unity, and then approaches the “full aging” exponent value of unity in the long time-limit. Consequently, any power‑law fit of our numerical predictions is inherently dependent on the chosen fitting window, as always also true in practical experimental or simulation studies. The second, much slower growth regime of the alpha time arises due to the slower component of the structural relaxation discussed above that occurs on a time scale comparable to quiescent alpha time, which is predicted to be beyond the timescale the stress has relaxed. The sharp change in aging rate arises because the external stress, acting as a microscopic force on individual particles, strongly softens the kinetic constraints and thus dominates the aging dynamics. Once the stress has fully relaxed, aging proceeds only via structural relaxation, resulting in a much slower aging rate.

The above rich behaviour arises directly from the time-dependent nonequilibrium rebuilding of the microscopically predicted activation barriers due to \textit{coupled} stress and structural relaxation. Figure \ref{fig3}(b) shows the time evolution of the local cage (solid), collective elastic (dashed), and total (dotted for $\phi=0.63$) barriers for different packing fractions. Initially, the barriers are at their low shear-melted steady state values. After shear cessation, they begin to rebuild as stress and structural deformation decrease, leading to the observed aging or re-equilibration behaviour. A key physical insight in high packing fraction systems is that the local cage barrier rebuilds faster than the elastic barrier, and it primarily governs the initial aging towards equilibrium. As discussed previously, this is a natural consequence within ECNLE theory of the latter involving longer length scale features of the dynamic free energy \cite{47mirigian2014elastically-4c6,74ghosh2020role-ee9,75mei2024mediumrange-ac8}. At longer times, when the local cage barriers have reached roughly $80\%$ of their quiescent value, the rebuilding of the collective elastic barriers begins to become significant, driving the final re-equilibration process towards recovery of the total barrier and quiescent relaxation time. The cage (elastic) barrier recovers more quickly (slowly) as packing fraction increases. This suggests that the degree of cooperativity of the mixed local-nonlocal alpha relaxation process of ECNLE theory, as quantified by the ratio of the elastic to local barrier (related to fragility \cite{76mei2020thermodynamicsstructuredynamics-ca8}), takes longer to develop at higher packing fractions. Remarkably, and nontrivially, although the two barriers show distinct temporal rebuilding behaviors, the total barrier that enters the alpha time [Eq.(\ref{eqn:TauAlpha})] is predicted to evolve in a rather simple manner. This reflects a fundamental causal connection between the two barriers in microscopic ECNLE theory since both are determined by the underlying dynamic free energy \cite{47mirigian2014elastically-4c6,75mei2024mediumrange-ac8}.

The rebuilding kinetics of the elastic shear modulus is shown in Fig.\ref{fig3}(c). The main panel shows its time evolution in a dimensionless format for different packing fractions, while the inset displays the unscaled results. During deformation, the reduction of $G^\prime$ is relatively small, resulting in a limited amplitude recovery response. A temporal two-step modulus build up is predicted with the initial increase driven by stress and structural recovery, and the final increase after stress relaxation due to relaxation of temporally frozen-in structure. For the highest packing fractions considered, the quiescent alpha relaxation time lies well beyond the considered time window, resulting in an effectively frozen‑in structural deformation and incomplete recovery of the elastic modulus. Full recovery is nevertheless expected at sufficiently long times, on the quiescent alpha time scale, thus leading to a sigmoidal two-step response. In practice, small amplitude oscillatory shear (SAOS) at fixed frequency can be performed at specified waiting times after cessation to probe the full aging description of the elastic shear modulus.
\subsection{Stress Relaxation}
\vspace{-0.3cm}
\begin{figure*}
    \centering
    \includegraphics[width=0.850\textwidth]{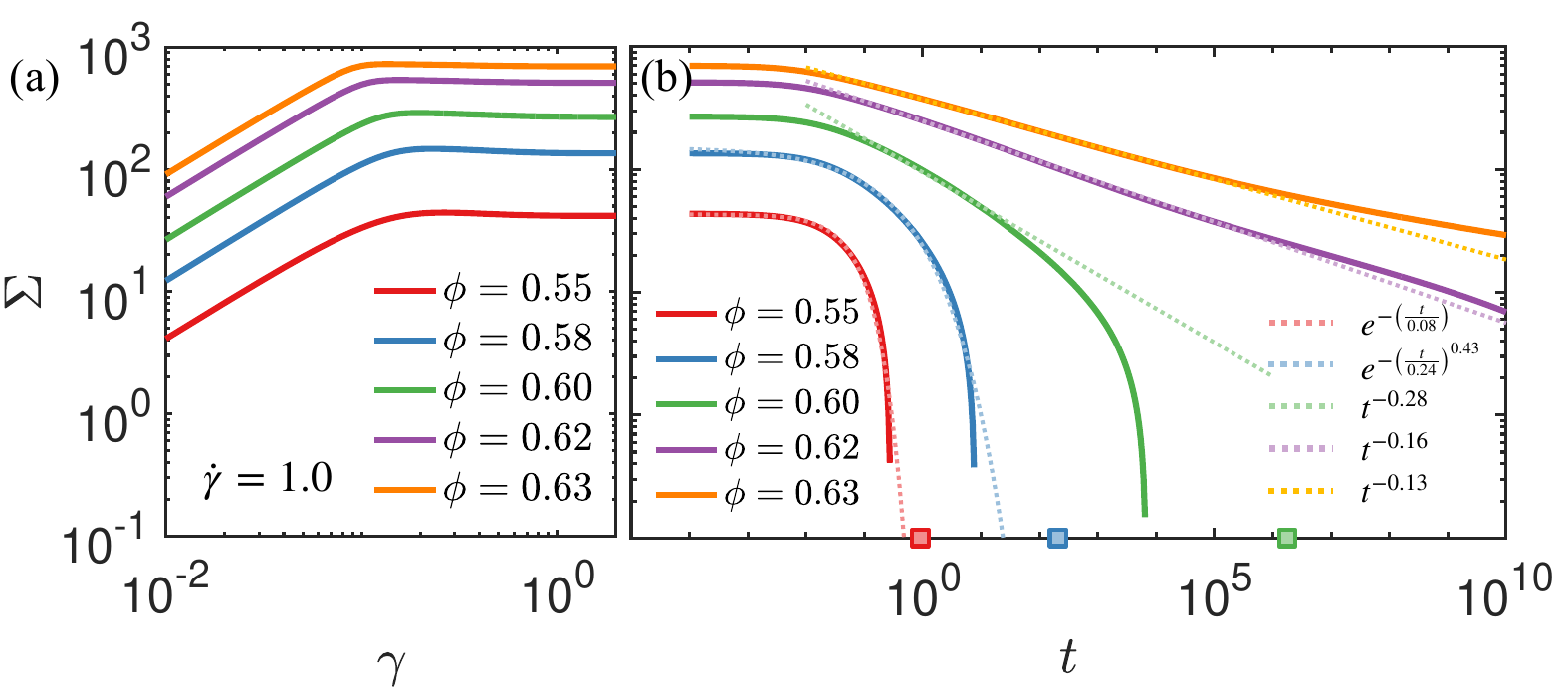}
    \caption{\textbf{Stress Relaxation from Steady-State Flow: Packing Fraction Dependence.} (a) Continuous start-up deformation stress-strain profiles at a constant shear rate of $\dot{\gamma}=1.0$ for different packing fractions. Overshoots are barely or not visible because of the use of an wide logarithmic scale. (b) Stress relaxation profiles following deformation cessation from the steady-states of panel (a). For lower packing fractions ($\phi\le0.55$), the stress relaxation follows a simple exponential decay, as indicated by the dotted curve. As the packing fraction increases, the relaxation profile evolves into a stretched exponential (KWW) form, with the stretching exponent dependent on $\phi$. For packing fractions $\phi\geq0.60$, power-law relaxation emerges, with a fractional exponent decreasing as $\phi$ increases (shown by the dotted curves). An inflection point is observed at larger times for $\phi\geq0.62$, where slower relaxation occurs due to the rebuilding of elastic barriers.}
    \label{fig4}
\end{figure*}
Figure \ref{fig4} presents the results for the temporal evolution of stress relaxation as a function of packing fraction. At smaller, but still deeply metastable, packing fractions ($\phi\approx0.55$), a simple \textit{exponential} stress relaxation process is predicted. The effective relaxation time is smaller than, but comparable to, the quiescent alpha time; for Model-B, it matches closely the quiescent relaxation time. As the packing fraction increases and the elastic barrier becomes important in equilibrium [Fig.\ref{fig1}(d)], the stress relaxation function transitions to a \textit{stretched exponential} form, with a KWW exponent of $\sim0.43$ for $\phi=0.58$. This is a direct consequence of the system losing memory of the deformation with time, leading to aging. The stress eventually vanishes on a time scale shorter than the quiescent alpha time. Further densification ($\phi\geq0.60$) extends the slow aging regime to longer timescales with an emergent slower \textit{apparent fractional power-law} relaxation, which eventually decays to zero exponentially at ultra-long times. Our comments above about the origin of the predicted apparent power laws again apply to these trends and behaviors. These fits of our numerical theoretical data are indicated in the figure, with fitting windows chosen to maximize the fit accuracy. The apparent power-law exponent decreases with packing fraction. The predicted very slow decays are indicative of the emergence of an apparent “residual stress” on experimental timescales. Given the Brownian nature of the model studied and the prediction of ECNLE theory of a non-inifinite barrier below RCP, the stress ultimately relaxes to zero on timescales comparable to the quiescent structural relaxation time. Thus, by “residual stress” we refer to the portion of stress that remains unrelaxed on experimentally accessible times, rather than a literal asymptotic long‑time stress plateau.

For the highest packing fractions considered, a transition to an even slower decay is observed after significant elapsed time, the functional form of which is difficult to unambiguously identify. Within the theory, they are associated with the system regaining elastic strength once the local cages are structurally re-established. Moreover, at such very high packing fractions the temporal rebuilding of the elastic barrier dominates over its cage analog at long times per Fig.\ref{fig3}(b). Hence, as an important point of broader significance in glass physics, this nonequilibrium behavior is a novel observable probe of the role of collective elasticity in the activated dynamics of glass forming matter, which is a core idea underlying the successes of ECNLE theory in equilibrium. However, this behavior should not be confused with the two‑step stress relaxation observed in soft‑jammed systems \cite{26mohan2013microscopic-571,27vasisht2022residual-eef,28vinutha2024memory-113} , where an initial rapid stress decay is associated with elastic cage recoil, followed by an extremely slow relaxation toward a stress plateau driven by structural re‑equilibration.

\begin{figure}
    \centering
    \includegraphics[width=0.490\textwidth]{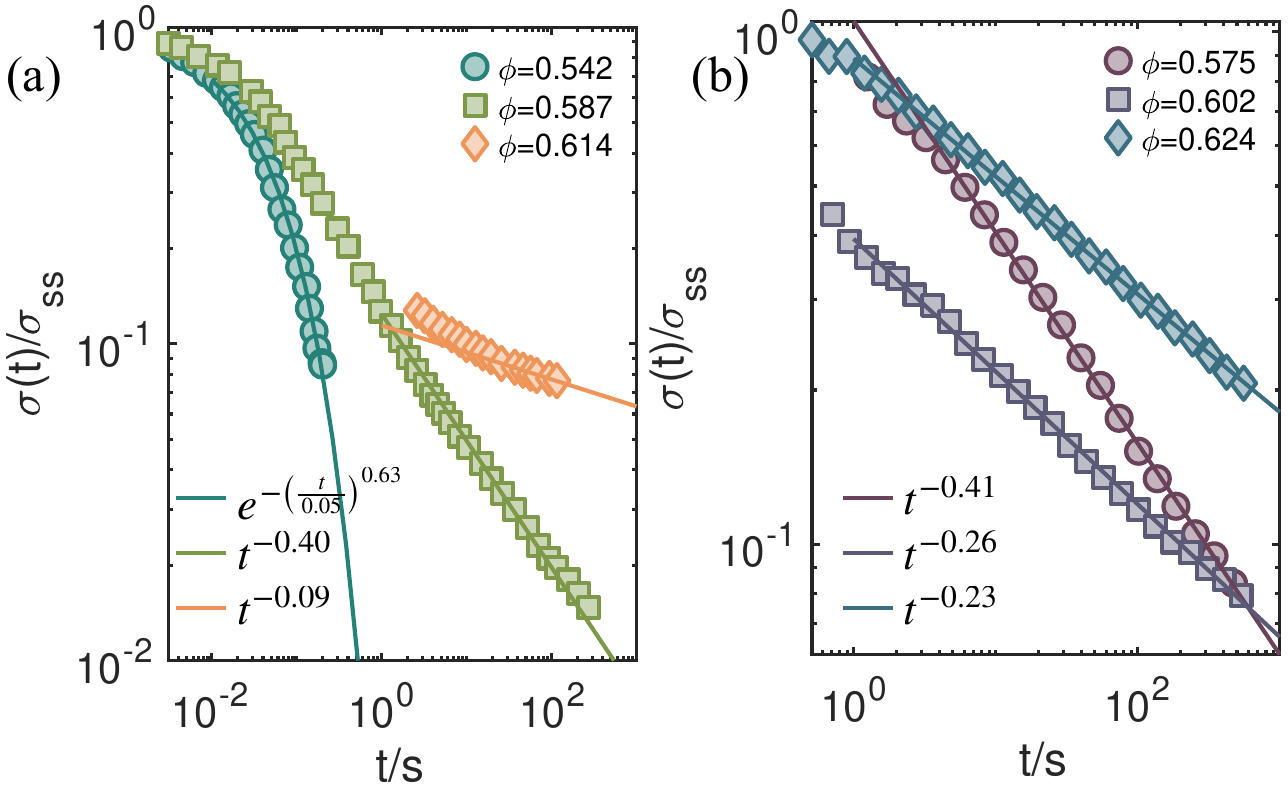}
    \caption{\textbf{Experimental Stress Relaxation Results.} Panel (a) presents the stress relaxation profiles extracted from Ref.\cite{20ballauff2013residual-748} following shear cessation from the steady state of a step-rate experiment for PMMA hard sphere colloid suspensions with $6\%$ polydispersity and a mean radius of $R=267~nm$ at the various indicated high packing fractions. The results exhibit a stretched exponential response at lower packing fractions, and a power-law decay with decreasing exponents at higher packing fractions. Panel (b) shows similar results extracted from Ref.\cite{23pamvouxoglou2021stress-937} for $25\%$ polydisperse PMMA particles with a mean radius of $R=~157 nm$. The stress relaxation profiles display a fractional power-law decay with decreasing exponents as packing fraction increases. These experimental results align well with the theory predictions for Model-A.}
    \label{fig5}
    \vspace{-0.3cm}
\end{figure}

The \textit{qualitatively} different predicted relaxation profiles—exponential, stretched-exponential (KWW), and fractional power-law decay—with growing packing fraction are consistent with both experiments \cite{17mckenna2009soft-b45,20ballauff2013residual-748,21jacob2019rheological-986,22chen2020microscopic-96e,23pamvouxoglou2021stress-937,30negi2010physical-17c} and simulations \cite{20ballauff2013residual-748,66marenne2017unsteady-1ac,77zausch2009buildup-5f3}. Moreover, and importantly, the so-observed KWW and fractional power-law decay exponents fall within the ranges observed here. For example, Ref. \cite{20ballauff2013residual-748}  measured the stress relaxation profile post steady state of step-rate experiments on PMMA hard sphere colloids with $6\%$ polydispersity and a mean radius $R=267nm$. The data is reproduced in Fig.\ref{fig5}(a) and it follows a stretched exponential profile with a KWW exponent of $\sim0.63$ for $\phi=0.542$, a power law decay with an exponent of $\sim-0.40$ for $\phi=0.587$, and a much smaller power law exponent of $\sim-0.09$ for $\phi=0.614$. The data from Ref. \cite{23pamvouxoglou2021stress-937}  for $25\%$ polydisperse PMMA suspensions with mean radius R=157nm\ is reproduced in Fig.\ref{fig5}(b). It exhibits fractional power law stress relaxation profiles with exponents of $-0.41$, $-0.26$, and $-0.23$ for $\phi=0.575,\ 0.602,$ and $0.624$ respectively. Similar experimental results can be found in Refs. \cite{17mckenna2009soft-b45,21jacob2019rheological-986,22chen2020microscopic-96e,30negi2010physical-17c}. Of course, one cannot expect perfect numerical agreement of the predicted exponents with any specific experiment given the theory is approximate, it ignores HI, and real systems are size polydisperse while we study a monodisperse smooth hard sphere model. Nevertheless, the magnitude of these experimental values and their evolution with packing fraction match well with the theoretical ones. An additional advantage of the theory relative to experiments and simulations is that it enables access to extremely long time scale dynamics, allowing the full stress‑relaxation process to be resolved.

We believe our above results represent a major success not been previously achieved, even \textit{qualitatively}, by any other microscopic theory. For example, the predictions of the ideal mode coupling theory \cite{20ballauff2013residual-748}  (which has no thermally activated relaxation events) do not qualitatively look like our results in Fig.\ref{fig4}. In this regard, we again emphasize that our above (and below) results are \textit{no adjustable parameter} microscopic predictions. Our goal is not to fit experimental data, since our theory is not a rheological model fitting tool with adjustable parameters. Rather, we aim to capture qualitatively and semi-quantitatively rich experimental behaviors from a microscopic foundation built on a deformation-evolving nonequilibrium dynamic free energy including activated relaxation.

Concerning stress relaxation predicted if we adopt Model-B, the timescale quantitatively shifts to longer values but the \textit{same} shape evolution as for Model-A is retained (see Fig.\ref{fig.S1}). Qualitatively, exponential, stretched exponential, and fractional power-law decay profiles are predicted as packing fraction increases, buttressing the robustness of the theory.

\subsection{Structural Recovery}
\vspace{-0.3cm}
In our approach, recovery of the deformed structure to equilibrium is fundamentally coupled to stress relaxation. It proceeds via an \textit{inverse} form of wavevector advection corresponding to the re-building of structural correlations. This process can thus be interpreted as advection backflow, and is quantified by the location of the first peak of the structure factor, $k^\ast\left(t\right)$. Of interest now is the relative change from equilibrium, $\Delta k^\ast\left(t\right)=k^\ast\left(t\right)-k^\ast$, as a function of time for different packing fractions. 
\begin{figure*}
    \centering
    \includegraphics[width=0.750\textwidth]{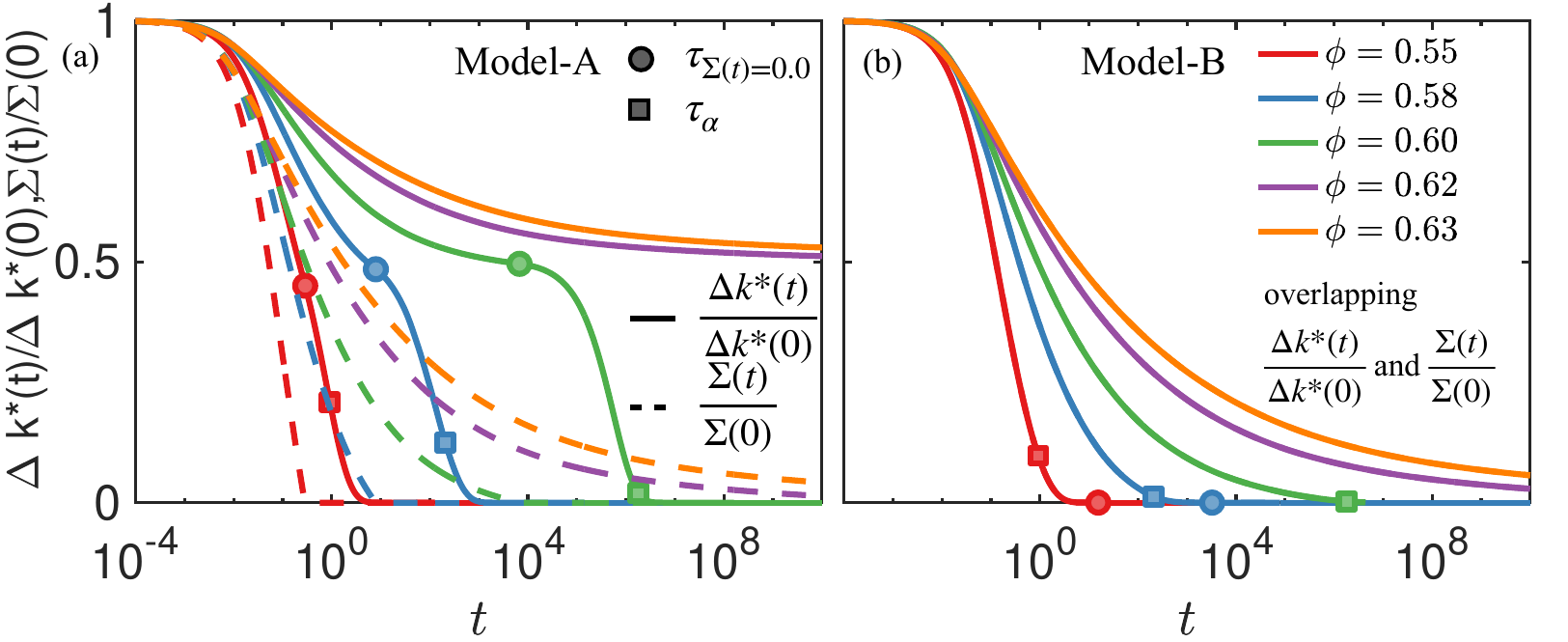}
    \caption{\textbf{Two vs One Step Structural Relaxation After Flow Cessation.} (a) The solid curves show the two-step structural recovery to equilibrium after flow cessation, with stress relaxation indicated by the dashed curves for different packing fractions, obtained using Model-A. Circles mark the time at which the stress reaches zero, while squares indicate the quiescent alpha relaxation time. (b) Same as (a), but obtained using Model-B, where the structure and stress evolution curves overlap. }
    \label{fig6}
\end{figure*}

Figure \ref{fig6} shows results in a normalized format for both Model-A and Model-B. Model-A  [panel (a)] includes additional stress reduction due to convective elastic backflow resulting in faster aging towards equilibrium, and a two-step structural relaxation that becomes more prominent with increasing packing fraction. In contrast, for Model-B [panel (b)], structural relaxation follows stress relaxation (with overlapping curves) and exhibits a $\phi$-dependent decay profile, which can be exponential, stretched exponential, or power law decay. The circles in the plots mark the time when the stress relaxes to zero in a practical experimental sense, which correlate well with the inflection point of the final structural relaxation event for Model-A. The squares represent the quiescent alpha times which align closely with the structural relaxation times of the non-equilibrium steady state in Model-A. Recall that in Model-A, the decoupling of structure and stress is driven by initially faster stress reduction compared to structural relaxation, which leads to faster aging, resulting in two-step structural relaxation. Interestingly, we note that this concept is invoked in granular suspensions \cite{25mohan2015buildup-d4d,26mohan2013microscopic-571,27vasisht2022residual-eef,28vinutha2024memory-113}, albeit with reversed roles: post-cessation, rapid structural rearrangements (compared to stress reduction) cause a rapid stress drop, followed by slower stress relaxation process, resulting in two-step stress relaxation. For extremely dense systems, we estimate the two-step structural relaxation can extend to extremely long timescales beyond the ability of practical laboratory experiments to probe, implying residual frozen structural deformation and stress, effectively imprinting the memory of the deformation on the structure.  
\vspace{-0.3cm}
\section{Pre-Shear Rate and Mechanical Memory Effects}
\vspace{-0.3cm}
Given the post-cessation recovery dynamics is initiated from the flowing nonequilibrium steady which is strongly affected by the applied shear rate, it is of high interest to explore what the present theory predicts for the influence of pre-shear deformation rate. The latter is a form of a nonequilibrium mechanical memory effect.

\subsection{Shear Thinning, Stress Relaxation, and Emergent Time Scaling}
\vspace{-0.3cm}
It is phenomenologically well established that the steady-state stress typically increases with shear rate as a fractional power law, the so-called empirical Herschel-Bulkley law \cite{78larson1998structure-407}, and the structural deformation relative to equilibrium is also shear rate dependent. Both of these effects have been successfully analyzed in our previous theoretical work for hard sphere suspensions \cite{45ghosh2023microscopic-39e}. Given the Brownian model studied and the physical ideas of ECNLE theory where the structural relaxation time never diverges below RCP for hard spheres, relaxation will always occur on a quiescent $\tau_\alpha$ timescale; thus, the steady-state stress vanishes in $(\dot{\gamma}\times\tau_\alpha)\rightarrow1$ limit, and the dynamic yield stress in the literal (typically not measureable)  zero-strain-rate limit is not predicted. 

\begin{figure}
    \centering
    \includegraphics[width=0.490\textwidth]{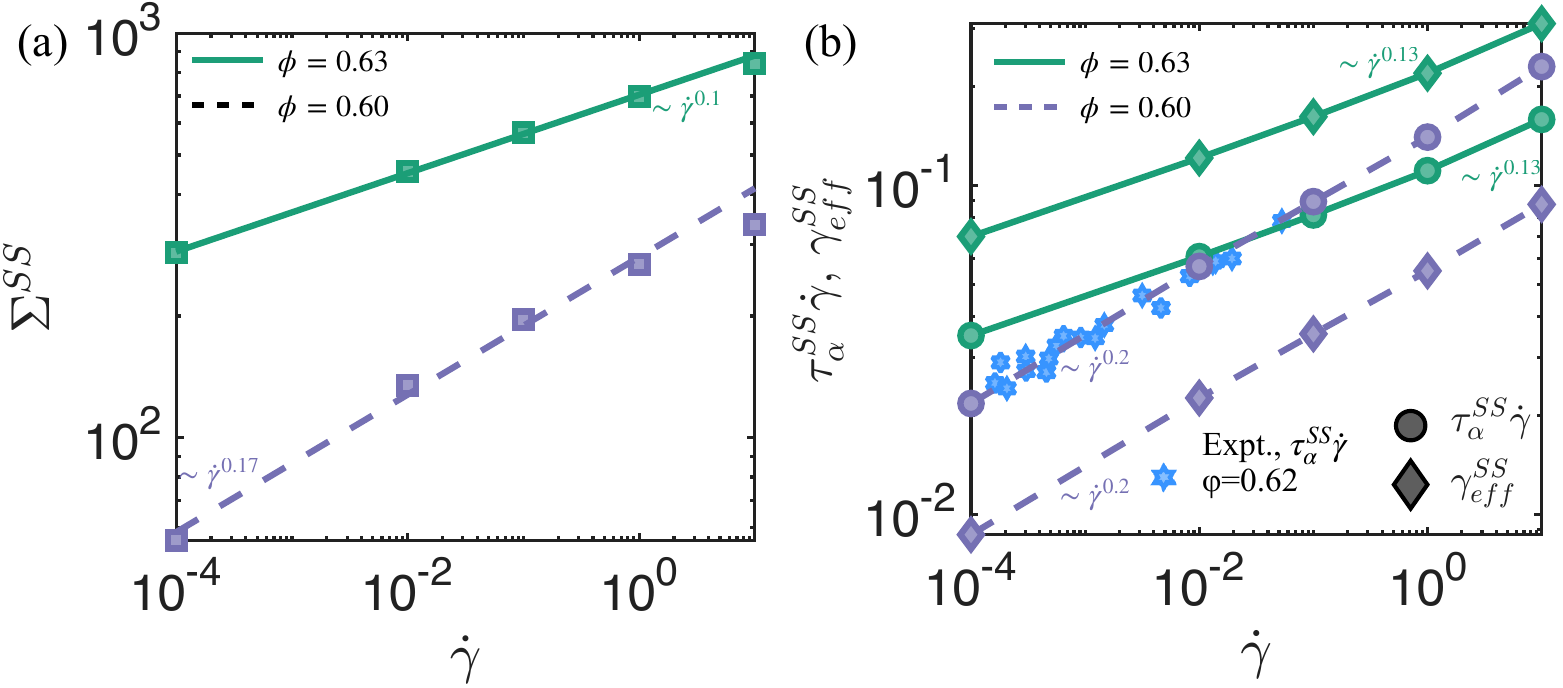}
    \caption{\textbf{Shear Thinning Response.} (a) Predicted steady state flow stress for continuous start-up shear experiments at two different high packing fractions as a function of shear rate. (b) Steady state shear thinning response of dimensionless effective Peclet number (circles), and steady state structural order parameter (diamonds). Apparent power law scalings are indicated. The blue stars indicates the experimental $\tau_\alpha^{SS}\dot{\gamma}\sim{\dot{\gamma}}^{0.2}$ behavior where the relaxation time was obtained from single particle trajectories and the incoherent dynamic structure factor under steady state shear.  This data is adapted from Ref.\cite{65besseling2007threedimensional-9cf}, and scaled by a constant numerical prefactor factor of $4.2$ to align the experimental and theoretical absolute timescales. See the text for system details.  }
    \label{fig7}
\end{figure}

The steady-state flow curves are illustrated in Fig.\ref{fig7} for $\phi=0.63$ (and $\phi=0.60$). The steady-state stress in Fig.\ref{fig7}(a) follows the shear thinning power laws $\mathrm{\Sigma}^{SS}\propto{\dot{\gamma}}^{0.1}$ ($\mathrm{\Sigma}^{SS}\propto{\dot{\gamma}}^{0.17}$ for $\phi=0.60$). The structural scalar parameter in Fig.\ref{fig7}(b) follows $\gamma_{eff}^{SS}\propto{\dot{\gamma}}^{0.13}$ ($\gamma_{eff}^{SS}\propto{\dot{\gamma}}^{0.20}$for $\phi=0.60$), while the reduced Peclet number follows  $\dot{\gamma}\tau_\alpha^{SS}\propto{\dot{\gamma}}^{0.13}$ $(\dot{\gamma}\tau_\alpha^{SS}\propto{\dot{\gamma}}^{0.20}$ for $\phi=0.60$). The shear thinning behaviour deviates from the ideal scaling $\dot{\gamma}\tau_\alpha^{SS}\sim$ constant for well understood physical reasons within the theory. The deviation becomes more pronounced at lower (but still high metastable) packing fractions, consistent with experiment and simulation \cite{56hebert2015effect-983,68koumakis2013complex-b0e,79koumakis2011study-b2a}. Moreover, consistent with the discussion above, various studies \cite{80eisenmann2009shear-e56,81jacob2015convective-227}  also find ideal shear thinning behaviour of $\tau_\alpha^{SS}\sim\dot{\gamma}$ for extremely dense metastable glassy systems. 

In Figure \ref{fig7} (b), we also show the shear thinning of the alpha time experimental data from Ref. \cite{65besseling2007threedimensional-9cf} obtained by tracking particle level trajectories for ultra-dense PMMA hard sphere colloid suspensions with $10\%$ polydispersity and $\phi\approx0.62$. These results relate \textit{directly} to our theoretical prediction of single particle relaxation times, and the experimental data follows the power-law relation $\dot{\gamma}\tau_\alpha\sim{\dot{\gamma}}^{0.2}$. The latter compares well with our predictions for monodisperse suspensions at $\phi=0.6$. These shear thinning behaviors are important for the present problem since they directly relate to how the initial stress relaxation after flow cessation depends on (pre) shear rate, which is germane to understanding the experimentally observed \cite{20ballauff2013residual-748}  time scaling by the inverse of the shear rate. Such scaling thus becomes increasingly effective for denser systems.   
\begin{figure*}
    \centering
    \includegraphics[width=0.950\textwidth]{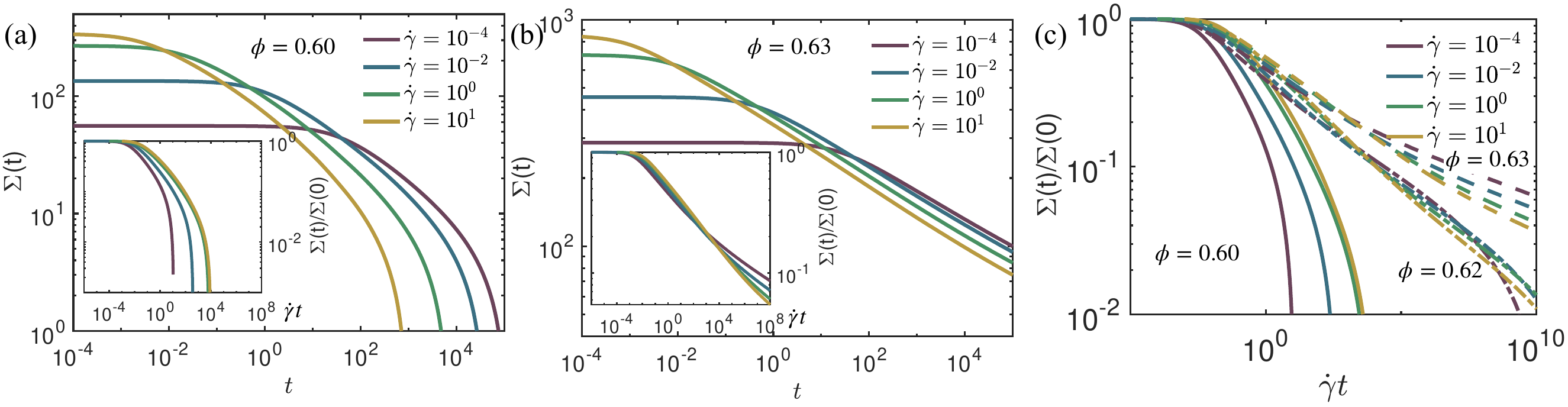}
    \caption{\textbf{Pre-Shear Strain Rate Dependence of Stress Relaxation from Steady-State Flow.} (a) Stress relaxation profiles for $\phi=0.60$, following the steady-state of continuous start-up shear at varying strain rates that span five decades. The stress vanishes within the shown time window, with higher pre-shear strain rates leading to earlier stress decay. (b) Stress relaxation profiles for $\phi=0.63$, showing a residual stress that decreases with increasing pre-shear strain rate. Insets in both panels display doubly reduced data, where stress is normalized by the steady-state stress and time is scaled by the inverse of the strain rate to achieve data collapse. (c) Doubly reduced stress relaxation profiles for 3 packing fractions (represented by different line styles) and 4 pre-shear strain rates (indicated by different colors). The two different sets of relaxation profiles ($\phi=0.60$ and $\phi=0.63$) are separated by a packing fraction of $\phi=0.62$, which marks an operationally defined glass transition density, as suggested in Ref.\cite{20ballauff2013residual-748}. }
    \label{fig8}
\end{figure*}

The question now is whether and/or how the stress or structural relaxation decays to either a finite non-zero value or to zero \textit{on} the experimental time scale for different pre-shear strain rates. For systems without residual stress, experiments show that stress relaxes more rapidly following higher pre‑shear rates, whereas in systems where stress does not fully relax, the long‑time relaxation profiles are parallel across different strain rates \cite{20ballauff2013residual-748}. Our present theory captures both these responses, as we now discuss.

Figures \ref{fig8}(a) and \ref{fig8}(b) present stress relaxation profiles for $\phi=0.60$ and $\phi=0.63$, respectively, after flow cessation for 4 shear rates spanning 5 decades in magnitude. For the $\phi=0.60$ system, the stress fully relaxes within the studied time window, with faster decay to zero for states prepared at higher shear rates. In contrast, for the denser  $\phi=0.63$ system, the residual stress is smaller for higher shear rates. These observations align with the predicted shear thinning behavior discussed above, where higher shear rates lead to larger initial stress relaxation and faster decay or smaller residual stress. Also, the stress relaxation curves for different strain rates are parallel, as experimentally observed \cite{20ballauff2013residual-748,23pamvouxoglou2021stress-937}. This suggests a simple time scaling occurs as a consequence of the predicted shear thinning behaviors. The analogous ideal MCT \cite{20ballauff2013residual-748}  curves are also parallel, but their shapes are entirely different from our results, and not align well with experiment.

In a search for simplicity, the insets of Figs. \ref{fig8}(a) and \ref{fig8}(b) show our results collapse rather well upon scaling the stress by the steady-state stress and time by the inverse shear rate. For the $\phi=0.60$ system, the profiles at higher shear rates collapse very well. For higher packing fractions, a near complete collapse is observed except at very long times where the memory of initial deformation is completely erased. Notably, we also observe an inversion of trends in the doubly normalized plots upon transitioning from lower to higher packing fractions. Specifically, for lower packing fractions, the higher shear rate curves remain above their lower shear rate analogs with no curve-crossing, while for larger packing fractions there is curve crossing. At short times, the higher shear rate profiles start above, but at longer times they end up below, the lower shear rate profiles. In the context of our results which are all for dense metastable states, “lower” and “higher” packing fractions connect back to the nature of the quiescent alpha process in ECNLE theory where the elastic barrier is unimportant until the dynamics becomes sufficiently slow at high enough packing fractions \cite{47mirigian2014elastically-4c6,75mei2024mediumrange-ac8,82mirigian2013unified-329}. Hence, we believe that the new nonequilibrium results in Figs. \ref{fig8}(a) and \ref{fig8}(b) shed further light on the fundamental question of the degree of spatial locality of the \textit{quiescent} activated glassy dynamics, and the core physical idea of two coupled barriers in ECNLE theory.

\subsection{Apparent Residual Stress and Locating the Glass Transition}
\vspace{-0.3cm}
Figure \ref{fig8}(c) shows the stress relaxation profiles for three packing fraction systems deformed at various pre-shear rates. The theory predicts at $\phi=0.62$ a clear boundary between two distinct behaviors emerges. For lower packing fractions, the stress relaxation curves for different shear rates do not collapse, suggesting a loss of memory. In contrast, for higher packing fractions the curves collapse, and the residual stress exhibits a shear rate dependence. At the crossover packing fraction $\phi=0.62$, the stress relaxation curves are nearly identical and collapse over the entire time range, and such a distinctive response has been hypothesized to be indicative of a “glass transition” \cite{20ballauff2013residual-748}. Moreover, they follow a remarkable apparent fractional power scaling of roughly $(\dot{\gamma}{t)}^{0.15}$. This exponent is sufficiently small that it might appear as a logarithmic decay in experiments or simulations.
\begin{figure*}
    \centering
    \includegraphics[width=0.750\textwidth]{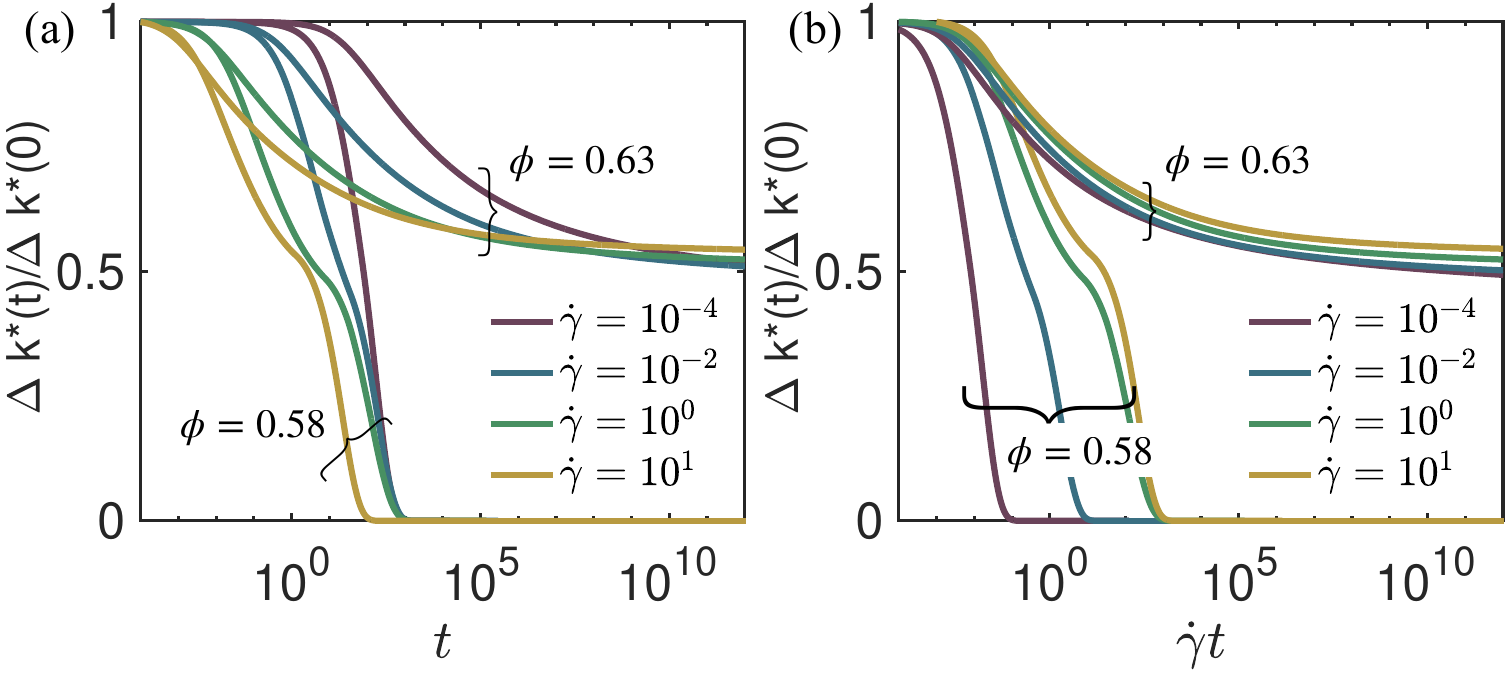}
    \caption{\textbf{Pre-shear Rate Dependence on Structure Relaxation.} (a) Time evolution of the structural order parameter for four different pre-shear rates and two high packing fractions. (b) Collapse of the time evolution plots upon rescaling time with strain rate.  }
    \label{fig9}
\end{figure*}
\subsection{Pre-shear Rate Dependence of Structural Relaxation}
\vspace{-0.3cm}

The pre-shear rate dependence discussed above is also evident in the structural relaxation profiles which are shown for Model-A in Fig.\ref{fig9}a for $\phi=0.63$ and $\phi=0.58$. At lower packing fractions, the structural relaxation changes from a single-step decay at smaller shear rates, to a two-step decay with faster relaxation at higher shear rates. This variation arises because the initial decay is governed by the shear rate-dependent non-equilibrium steady state. In contrast, the final structural relaxation occurs at a time close to the quiescent alpha time leading to a single-step decay for smaller shear rate, where the onset time is large and comparable to quiescent alpha time. For higher shear rates, the onset times are shorter and well separated from the quiescent alpha time, resulting in a two-step decay. Consequently, higher shear rates lead to faster relaxation, but retain deformation memory for a longer duration, in contrast to smaller shear rates that relax primarily through the quiescent process. At higher packing fractions, only the initial decay to a plateau is observed, which is pre-shear rate dependent. 

Figure \ref{fig9}(b) shows the different curves in Fig.\ref{fig9}(a) are predicted to collapse upon non-dimensionalization of time by the shear rate applied during the start up preparation deformation before cessation. For lower packing fractions, the higher shear rate curves collapse, per the above discussion. In contrast, all curves collapse for higher packing fractions due to the well-separated time scales. Overall, the stress and structural relaxation profiles following shear cessation for different pre-shear rates reveal key insights, including trapped residual stresses and structural deformations at relevant time scales. These profiles are closely tied to the predicted aging behaviour after deformation.

\section{Flow Cessation at Variable Strain Loadings}\label{Sec7}
\begin{figure*}
    \centering
    \includegraphics[width=0.750\textwidth]{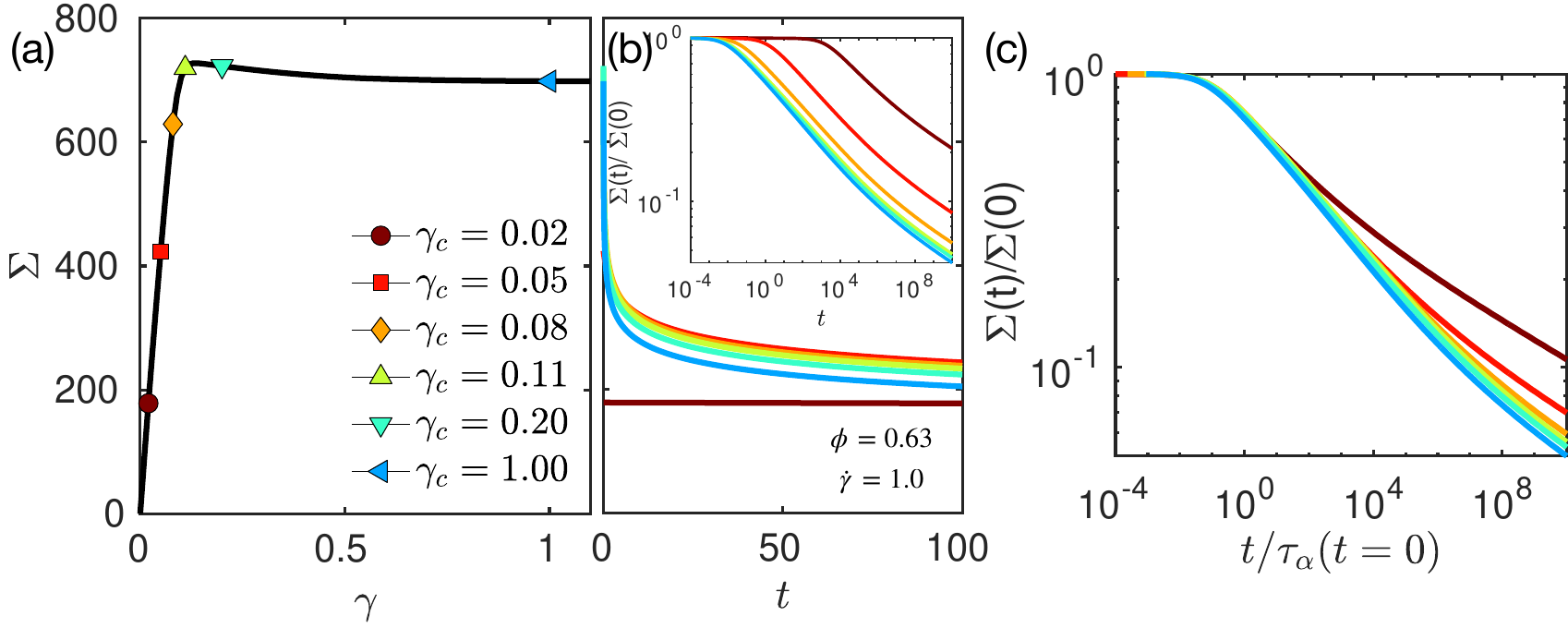}
    \caption{\textbf{Stress Relaxation Profile After Flow Cessation at Different Strains} (a) Stress-strain curve at $\phi=0.63$ subjected to continuous start-up shear with $\dot{\gamma}=1.0$. Points indicate the different strain values at which flow cessation is applied. (b) Stress relaxation profiles for Model-A displayed on a linear scale, with the inset showing the scaled stress relaxation (normalized by the initial stress) in a logarithmic representation. (c) Collapse of the stress relaxation curves achieved by rescaling time with the initial shear-reduced $\alpha$-time corresponding to the selected strain cessation point.}
    \label{fig10}
\end{figure*}

\vspace{-0.3cm}
We have briefly examined the stress relaxation response initiated from different cessation points in a continuous start-up shear experiment before the nonequilibrium steady state was reached.  Fig.\ref{fig10}(a) shows the stress-strain curve for $\phi=0.63$ and a shear rate of $\dot{\gamma}=1.0$. The different points on the plot correspond to selected cessation points, from which the stress relaxation profiles are obtained. Panel (b) displays the stress relaxation profiles on a linear scale over a short time window, color-coded according to the cessation points shown in panel (a); the inset provides the full relaxation profile on a log-log scale. 
Several key observations can be made. (i) For stress cessation in the elastic regime, the response remains almost unchanged at longer times after the initial incubation/onset time, as can be observed from almost parallel long time stress relaxation profiles shown in the inset of Fig.\ref{fig10}(b). This onset time sensibly decreases as the cessation strain increases. (ii) The relaxation profiles for cessation at strains near the stress overshoot and in the steady state are nearly identical, suggesting a similarity of these non-equilibrium states. This buttresses the interpretation of the stress overshoot as a crossover from an elastic-like to a viscous-like mechanical response. (iii) Overall, the stress relaxation profiles are similar, differing mainly in their onset times. In panel (c), we scale the elapsed time by the alpha relaxation time at each cessation point, which results in an interesting collapse of all the relaxation profiles. We suggest that the quality of the collapse warrants the use of stress relaxation profiles as an alternative method to study shear thinning. 

\begin{figure*}
    \centering
    \includegraphics[width=0.750\textwidth]{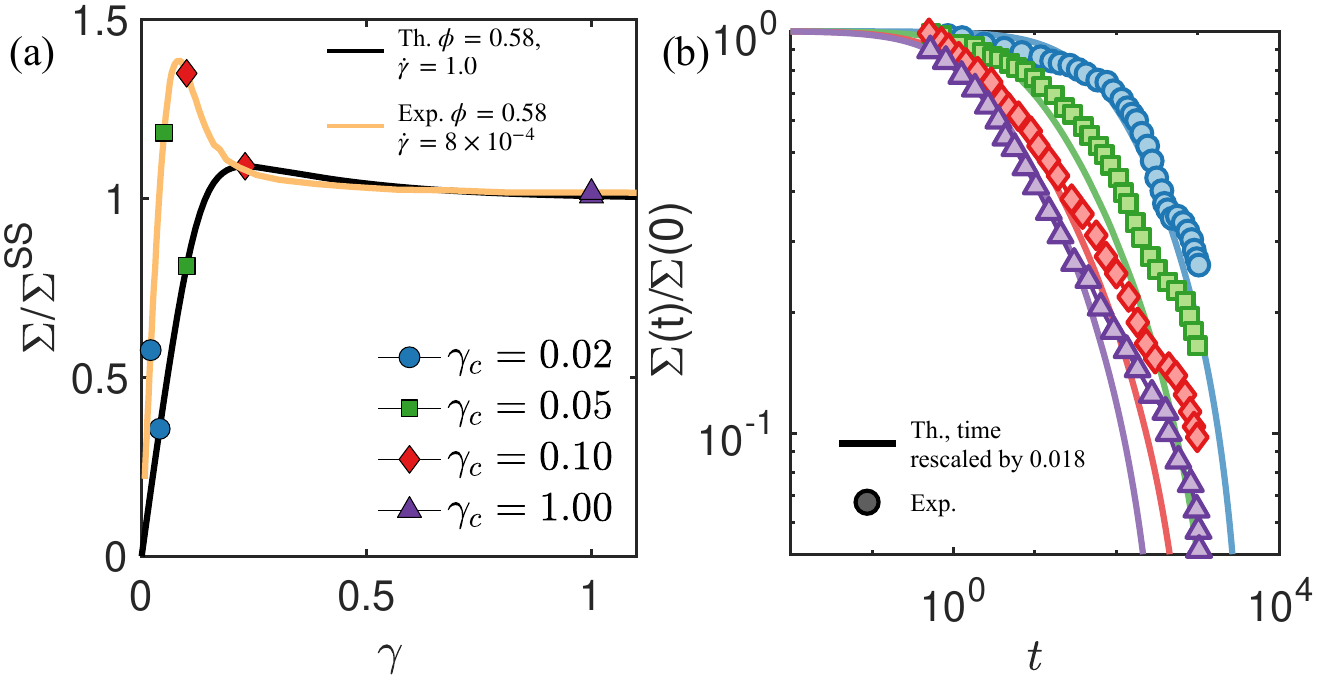}
    \caption{\textbf{Comparison of Shear Cessation at Different Strains with Experiment.} (a) Theory predictions for the continuous start up shear response at a packing fraction of $\phi=0.58$ and experimental data from a polydisperse dense colloidal suspension with particle radius $R=106\ nm$ suspended in squalene at $\phi=0.58$\cite{21jacob2019rheological-986}. (b) Stress relaxation curves at various qualitatively similar marked points from (a) are shown for both the theoretical (Model-A) and experimental systems.}
    \label{fig11}
\end{figure*}

We have also attempted to \textit{qualitatively} compare our theoretical stress relaxation profiles from different cessation points to the experimental results of Ref. \cite{21jacob2019rheological-986}  for polydisperse dense colloidal suspensions with particle radius $R=106\ nm$ suspended in squalene at $\phi=0.58$.  Figure \ref{fig11}(a) shows the theory predictions of step-rate shear response of monodisperse hard spheres of $\phi=0.58$, along with the experimental results. The absolute magnitude of the overshoot stress and strain differs for experiment and theory as expected, and \textit{no attempt} is made to quantitatively align or fit them. Rather, our goal is to qualitatively compare the theory and experimental stress relaxation responses from elastic, pre-overshoot, overshoot, and steady state strain cessation points, as marked by the different colored points. 

In panel (b), the theoretical stress relaxation curves are compared to experimental data at various similar cessation points. The time scale of the theoretical profiles have been rescaled by a factor of $0.018$ to allow a comparison of the shape of the decay curves. This quantitative rescaling may reflect multiple differences between experiment and theory including solvent viscosity, size polydispersity, HI, and colloid surface roughness. Overall, we find the relaxation profiles from different loading points exhibit good qualitative agreement between theory and experiment.  
\section{Summary and Future Outlook}
\vspace{-0.3cm}
We have developed a microscopic statistical mechanical theory that self-consistently predicts stress and structural relaxation and physical aging (or re-equilibration dynamics) after shear cessation from a mechanically prepared nonequilibrium state from knowledge of how microscopic force constraints, the elastic modulus, and activated structural relaxation time evolve under active deformation nonequilibrium conditions. The fundamental input that renders the approach microscopic and predictive is solely the deformed pair structure.

 Especially notable new advances include the following. (i) The theory correctly captures the evolution of the stress relaxation profiles with increasing particle loading, transitioning from exponential to stretched exponential to power-law decays with decreasing fractional exponents. (ii) A physically grounded description of aging following flow cessation is constructed which is fully coupled with stress and structural relaxation. (iii) Based on our modeling of elastic convective backflow (Model-A), a 2-step relaxation of structure is predicted in contrast to stress relaxation. Importantly, the shear-melted nonequilibrium steady state, regardless of whether the underlying quiescent system is a metastable liquid or a glass, exhibits very short relaxation times. This renders important taking into account activated dynamics far from equilibrium, as we have done in a predictive microscopic manner within the nonequilibrium version of ECNLE theory. (iv) The spatially nonlocal nature of the alpha relaxation process predicted by ECNLE theory under quiescent conditions involving coupled local cage and collective elastic barriers has crucial consequences for the rich nonequilibrium phenomena studied. This provides a new lens on probing the role of the proposed spatially nonlocal aspect of quiescent activated structural relaxation in glass-forming matter. 
 
The predicted stress relaxation response and aging behaviors are found to be in good agreement with diverse experimental observations. Notably, the near-ideal aging behavior, characterized by a power-law increase in the alpha relaxation time with an exponent close to unity following shear cessation, has been widely observed \cite{4joshi2018yield-302,17mckenna2009soft-b45,22chen2020microscopic-96e,73peng2014comparison-631,83bandyopadhyay2006slow-4b5,84joshi2008aging-676}. For instance, Ref. \cite{22chen2020microscopic-96e} reports that the relaxation times of the intensity autocorrelation functions, measured using x-ray photon correlation spectroscopy (XPCS) for charge-stabilized silica nanospheres in water, grow as $t^{1.12}$ after cessation of shear for strain amplitudes ranging from $2\%$ to $20\%$. Similarly, aging exponents between $0.4$ and $0.8$ were reported in Refs.\cite{17mckenna2009soft-b45,73peng2014comparison-631} for various soft matter systems undergoing creep after a shear-melting protocol. The predicted evolution of stress relaxation profiles with particle loading—transitioning from exponential to stretched exponential to power-law decay—is consistent with experimental findings in Refs.\cite{17mckenna2009soft-b45,20ballauff2013residual-748,22chen2020microscopic-96e,23pamvouxoglou2021stress-937,85song2022microscopic-68c}. Many studies report distinct parallel stress relaxation curves corresponding to different pre-shear rates, which supports the simple time-rescaling predicted, along with its suggested connection to shear-thinning in the precursor flowing state due to memory effects. 

The theory based on Model-A accounts for a simple form of convective elastic backflow effects on nonequilibrium structural evolution, and qualitatively captures the experimentally observed \cite{17mckenna2009soft-b45,36denisov2013resolving-853} decoupling between stress and structural relaxation. More generally, this formulation predicts elastic stress reduction during recovery results in faster stress relaxation and a two-step slower structural relaxation process. Such a timescale mismatch has been observed experimentally in Ref. 36 where x-ray scattering measurements revealed that structural relaxation was slower than the corresponding mechanical stress relaxation. Similarly, Ref.\cite{17mckenna2009soft-b45} reported a structural relaxation time approximately 10 times longer than the mechanical relaxation time in aging measurements of latex particles ($50\%$ packing fraction) suspended in water, with increased decorrelation for larger packing fractions. On the other hand, Model-B ignores the physics of elastic effects associated with convective backflow and predicts a slaved one-step stress and structural relaxation. 

While structural and stress relaxation may occur on different timescales, in the calculations presented in the main article we adopted the minimalist assumption that the quantitative mismatch factor $f=1$ in Eq. 3. However, experimental observations \cite{17mckenna2009soft-b45,36denisov2013resolving-853}  and physical intuition suggest that structural relaxation could be slower than stress relaxation, as captured in our approach if $f<1$. Calculations based on the latter choice are briefly explored in the SI (Fig. S2). While the stress relaxation profiles remain largely unaffected by such variations in $f$, the structural relaxation can be systematically slowed. Hence, $f$ is a potentially system-dependent nonuniversal parameter which may be relevant to extensions of the theory to other soft matter materials and thermal glass-forming liquids.

We note that our analysis has ignored the near and far field interparticle hydrodynamic interactions, which we believe will not significantly affect stress and structure re-equilibration dynamics at the high packing fractions we focus on which are controlled to leading order by slower deformation-modified thermally activated processes. On the other hand, at lower packing fractions \textit{below} when activated dynamics become of paramount importance (not the focus of the present study), such HI effects can be more important and warrant further study.

As a future outlook, we emphasize that the new microscopic theoretical approach is quite general in that it can be readily applied to other diverse repulsive soft matter systems that interact via different potentials such as neutral and charged soft colloids, microgels, and emulsions. Furthermore, the qualitatively new effects associated with the introduction of short range inter-particle attractions which can induce physical bonding that competes with steric caging can also be treated. This includes systems that undergo double yielding under continuous startup shear deformation \cite{46mutneja2025microscopic-f31}  which are expected to display qualitatively new stress, structural and aging behaviors after flow cessation. Work in some these exciting future directions has been initiated.


%
%

%

\begin{acknowledgments}
The authors acknowledge support from the Army Research Office via a MURI grant with Contract No. W911NF-21-0146.
\end{acknowledgments}
\appendix
\section{Background quiescent theory}\label{Apd:A}
\vspace{-0.3cm}
\subsection{Activated Dynamics NLE Theory Basics}\label{Apd:A1}
\vspace{-0.3cm}
The starting point for describing quiescent fluid, activated single particle dynamics at the stochastic trajectory level is the overdamped, force balance, nonlinear Langevin equation (NLE)\cite{48schweizer2005derivation-c99} for the angularly averaged scalar displacement of a tagged particle from its initial position, $r(t)$: 
\begin{equation}\label{eqn:NLE}
    -\zeta_s\frac{dr\left(t\right)}{dt}-\frac{\partial F_{dyn}(r\left(t\right))}{\partial r(t)}+\delta f(t)=0
\end{equation}
The first term represents the dissipative non-activated short time and distance frictional drag force associated with “in cage” dynamics which is characterized by a short time friction constant $\zeta_s$, while the last term is the corresponding fluctuating white noise random force that obeys, $\left\langle\delta f\left(0\right)\delta f\left(t\right)\right\rangle=2k_BT\zeta_s\delta\left(t\right)$. This short time process defines the elementary diffusive unit of time in all our analysis and figures, $\tau_s=\beta\zeta_s\sigma^2$, which grows relatively weakly with increasing packing fraction for hard spheres; the explicit expression is (see Ref. \cite{48schweizer2005derivation-c99} ). 
\begin{equation}\label{eqn:Taus}
\begin{split}
    \tau_s\equiv\ \tau_0g(\sigma)\left[1+\frac{\sigma^3}{36\pi\phi}\int_{0}^{\infty}{dk\frac{k^2{(S(k\sigma)-1)}^2}{S(k\sigma)+b(k\sigma)}}\right],\\ b^{-1}(\sigma k)\equiv1-j_0(k\sigma)+2j_2(k\sigma)
    \end{split}
\end{equation}
Here, $j_n(x)$ is the spherical Bessel function of order $n$, the hard sphere fluid contact value of the pair correlation function is $g(\sigma)$, and $\tau_0\equiv\frac{\zeta_{SE}\sigma^2}{k_BT}$ is the most elementary “bare” time scale in suspensions written in terms of the Stokes-Einstein friction constant, $\zeta_{SE}$. The second crucial term in Eq (A1) is the effective force on any moving particle exerted by the other moving particles, and is defined as the negative gradient of the spatially-resolved microscopic dynamic free energy $F_{dyn}(r)$ (Eq. (1)). Activated motion emerges for hard spheres beyond the ideal näive (single particle) mode coupling theory (NMCT) kinetic “transition” at $\phi_c=0.44$ \cite{50zhou2020integral-c4b} which represents a dynamic crossover beyond which the dynamic free energy takes on a spatially localized form. 
\subsection{Dynamic Localization Length and Elastic Shear Modulus}\label{Apd:A2}
\vspace{-0.3cm}
The dynamic localization length in NMCT, $r_L$, or equivalently the minimum of the dynamic free energy \cite{47mirigian2014elastically-4c6,48schweizer2005derivation-c99} obeys the self-consistent equation:
\begin{equation}\label{eqn:RL}
    \frac{1}{r_L^2}=\frac{1}{9}\int{\frac{\vec{dk}}{\left(2\pi\right)^3}k^2\rho C^2(k)S\left(k\right)e^{-\frac{k^2r_L^2}{6}\left[1+S^{-1}\left(k\right)\right]}}
\end{equation}
Based on the classic idea that slow density fluctuations control slow stress dynamical fluctuations and emergent shear rigidity, the elastic shear modulus beyond the NMCT crossover is given by the well-known expression \cite{51nagele1998linear-c57},
\begin{equation}\label{eqn:GPrime}
    G^\prime=\frac{k_BT}{60\pi^2}\mathrm{\int_{0}^{\infty}}dk\left[k^2\rho S\left(k\right)\frac{dC(k)}{dk}\right]^2\exp{\left(-\frac{k^2r_L^2}{3S\left(k\right)}\right)}.
\end{equation}
\subsection{Activated Structural Relaxation Time in ECNLE Theory}\label{Apd:A3}
\vspace{-0.3cm}
As sketched in the main text, Kramers theory is employed to compute the mean first passage time for barrier crossing which in ECNLE theory includes the beyond cage scale collective elastic effects required to achieve a large amplitude local hopping event\cite{47mirigian2014elastically-4c6}. The elastic barrier is calculated within an Einstein glass description of the dynamical state of localized particles outside the cage as $F_{el}=4\pi\int_{r_{cage}}^{\infty}r^2\rho g\left(r\right)\left(\frac{1}{2}K_0u\left(r\right)^2\right)dr$, where $K_0$ is the harmonic spring constant of $F_{dyn}(r)$ at its minima at $r=r_L$, and $u\left(r\right)$ is the displacement field for particles outside the cage at a distance $r$ from the cage center. The elastic displacements are predicted to be small, and hence the displacement field is constructed in the spirit of continuum linear elasticity \cite{52dyre2006elastic-c3c}  as $u\left(r\right)=\Delta r_{eff}\left(\frac{r_{cage}}{r}\right)^2$ for $r\geq r_{cage}$, where the cage radius $r_{cage}$ is identified as the distance at the first minimum of $g\left(r\right)$. The displacement field amplitude ($\Delta r_{eff}$) is associated with the effective cage expansion length scale which \textit{a priori} follows from a microscopic analysis of the mean angularly-averaged extent to which hopping of all particles in the cage result in a displacement larger than the cage size. Defining the microscopic jump distance $\Delta r=r_B-r_L$, this analysis yields \cite{47mirigian2014elastically-4c6}
\begin{equation}
\Delta r_{eff}\approx\frac{3}{r_{cage}^3}\left(\frac{r_{cage}^2\Delta r^2}{32}-\frac{r_{cage}\Delta r^3}{192}+\frac{\Delta r^4}{3072}\right)			\end{equation}
Given the barriers, the mean alpha time follows from the Kramers theory as \cite{47mirigian2014elastically-4c6,53kramers1940brownian-150} 
\begin{equation}\label{eqn:TauAlpha}
    \frac{\tau_\alpha}{\tau_s}=e^{\beta F_{el}}\int_{r_L}^{r_B}{dx\ e^{\beta F_{dyn}(x)}}\int_{r_L}^{x}{dy\ e^{-\beta F_{dyn}(y)}}		
\end{equation}
Beyond a low total barrier of order 1-2 thermal energy units, the simpler and more intuitive form of Kramers theory applies
\begin{equation}
\frac{\tau_\alpha}{\tau_s}\approx\frac{2\pi}{\sqrt{K_0K_B}}\exp[\beta(F_B+F_{el})]				
\end{equation}
\begin{figure*}
    \centering
    \includegraphics[width=0.750\textwidth]{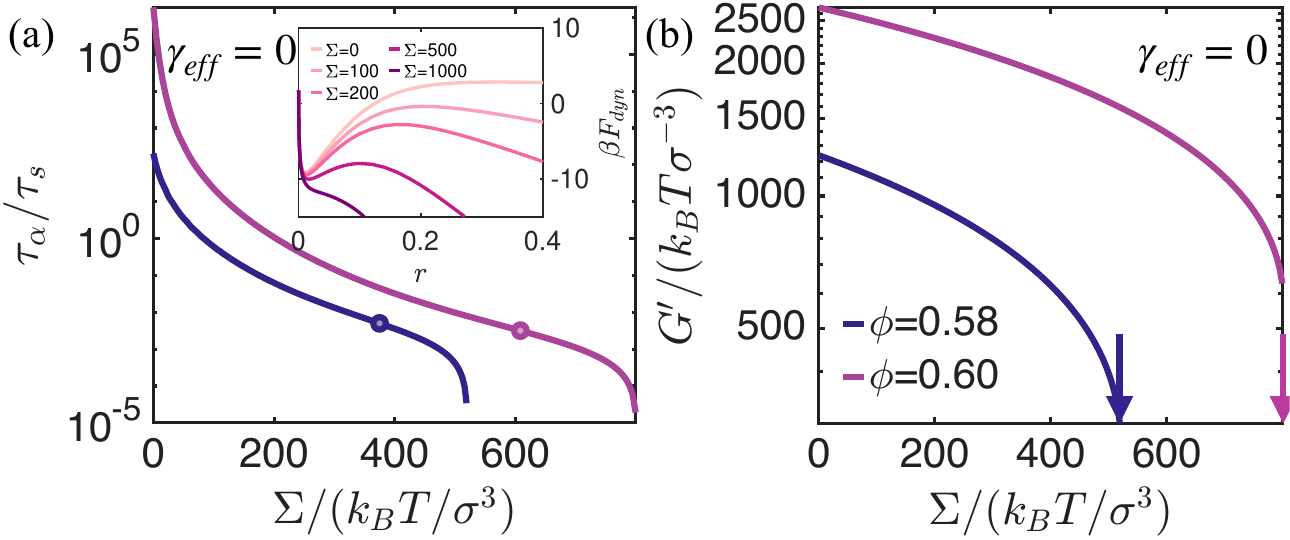}
    \caption{\textbf{Predicted effect of an imposed external stress.} (a) The inset shows the evolution of  the dynamic free energy with increasing external stress (or force). The main panel shows the non-dimensionalized alpha time evolution as a function of imposed stress for two different packing fractions. The points mark when the total barrier equals the thermal energy, indicating practical fluidization and a crossover from activated to non-activated particle trajectories. (b) The corresponding dimensionless elastic shear modulus evolution. The arrows mark the “absolute yield stress” at which the total barrier and localized state vanish and the dynamics is non-activated.  }
    \label{fig12}
\end{figure*}
Here, per Fig.\ref{fig1}(b) ,$K_0$ and $K_B$ are the absolute values of the harmonic curvatures of dynamic free energy at its minimum (the localization length $r_L$) and maximum (barrier location $r_B$), respectively.

\section{Integration of external stress into ECNLE theory}\label{Apd:B}
\vspace{-0.3cm}
Imagine a system is subjected to a macroscopic stress, $\Sigma$. As sketched in the main text, its consequences enter the NLE evolution equation as an effective additional microscopic force  $f_{ext}=A\mathrm{\Sigma}$ , with $A=\frac{\pi\sigma^2}{24}$ the relevant cross-sectional area \cite{60kobelev2005strain-2be}. The strongly anharmonic dynamic free energy thus acquires a nonequilibrium contribution of a mechanical work form: \cite{60kobelev2005strain-2be,74ghosh2020role-ee9} $\beta F_{dyn}\left(r,\mathrm{\Sigma}\right)=\beta F_{dyn}\left(r,\mathrm{\Sigma}=0\right)-f_{ext}r$. Increasing stress (even if changes of structure are ignored) reduces \textit{all} measures of transient solidity. For example, stress reduces the degree of particle localization and hence $G^\prime$, and the activation barrier and hence alpha relaxation time [Fig.\ref{fig12}(a) inset]. The parametric effect of changing external stress (with no change of structure) on the $\alpha$-time and elastic modulus is illustrated in Figs.\ref{fig12}(a) and \ref{fig12}(b), respectively. 

\section{Integration of structural deformation into ECNLE theory}\label{Apd:C}
\vspace{-0.3cm}
\begin{figure*}
    \centering
    \includegraphics[width=0.750\textwidth]{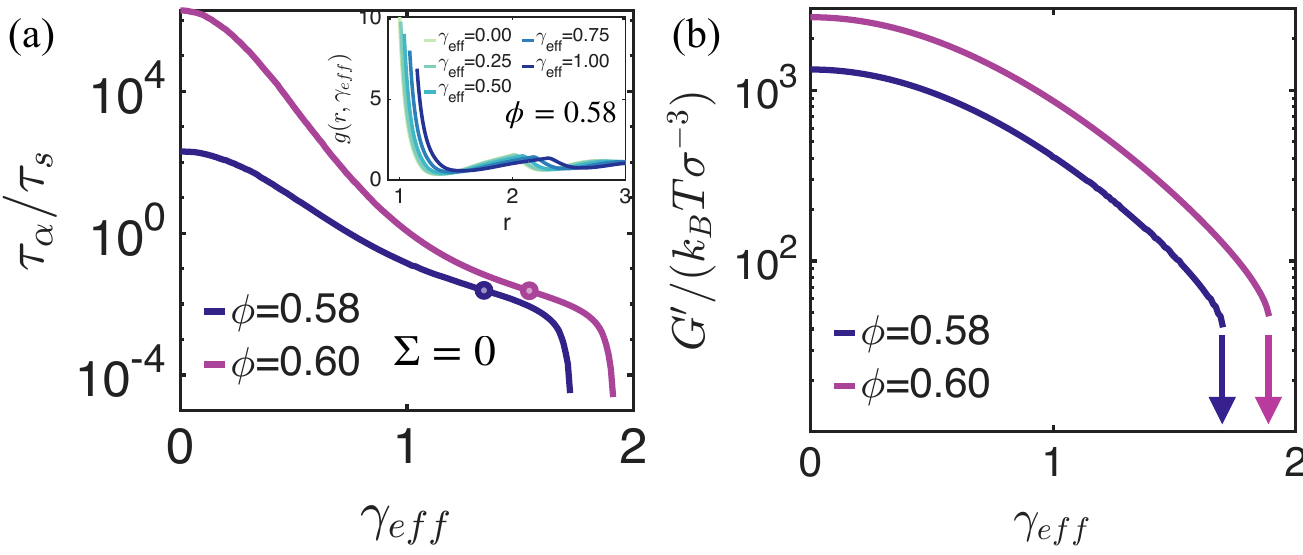}
    \caption{\textbf{Predicted effect of an imposed affine structural deformation} (a) The inset shows the pair correlation function evolution with affine structural deformation via wavevector advection. The main panel contains the non-dimensionalized alpha time evolution for two indicated packing fractions. The circles indicate when the total barrier (cage plus elastic) attains a value of $k_BT$, a practical limit beyond which an activated picture of particle motion no longer applies. (b) The corresponding elastic shear modulus evolution. The arrows mark the strain at “absolute yield” corresponding to a strain-induced fluidization transition at which the localized state and barrier of the dynamic free energy discontinuously and continuously vanish, respectively.}
    \label{fig13}
\end{figure*}
Structural deformation at a \textit{prescribed} effective microscopic strain $\gamma_{eff}$ is modeled in a no adjustable parameter manner using the wavevector advection concept. While the time-dependent evolution of $\gamma_{eff}$ for a real step-rate deformation is complex, it equals the bare rheological strain $\gamma$ at small deformation and saturates in steady flow. Figure \ref{fig13} explores the \textit{sole} effects of strain‑induced structural modification. This is a hypothetical parametric calculation with stress and hence microscopic external force is set to zero (no viscous relaxation events) in the dynamic free energy. 

The inset of panel (a) illustrates how the pair correlation function parametrically changes with $\gamma_{eff}$. The near contact region of $g(r)$ is increasingly suppressed corresponding to local structure disordering. This reduces the elementary interparticle collision rate for hard spheres, and hence all consequences of dynamic caging. The consequences of this structural softening \textit{alone} on the alpha relaxation time and elastic shear modulus are illustrated in panels (a) and (b), respectively. The relaxation time is strongly suppressed with increasing strain in a roughly 2-regime like manner. This dynamical softening initially begins for the collective elastic barrier which is sensitive to larger length scale features in the dynamic free energy, and then at larger strains for the local cage barrier \cite{74ghosh2020role-ee9}. Ultimately, when the total barrier becomes smaller than the thermal energy, a final stage of relaxation time acceleration occurs, corresponding to the barrier vanishing (the visually sharp drop regime), which is predicted to occur at what we call the “absolute yield strain”, where particles become delocalized and motion non-activated. Analogous behavior of the elastic shear modulus is seen in panel (b), but here the drop of $G\prime$ with strain is smoother since it is controlled by the smallest dynamical length scale, the localization length associated with the minimum of the dynamic free energy. The differing shape of the responses of the alpha time and elastic modulus reflects the fundamentally \textit{microscopic} and \textit{spatially-resolved} nature of the particle level based theory, in qualitative contrast with phenomenological coarse grained “trap models” \cite{40bouchaud1992weak-411,41monthus1999models-c48}. 

\bibliography{StressRelax}
\newpage
\setcounter{figure}{0}
\renewcommand{\thefigure}{S\arabic{figure}}
\onecolumngrid
\section*{Supplementary Information}
Additional figures that further extend and/or support the prime conclusions drawn in the main text are presented.
\begin{figure}[h]
    \centering
    \includegraphics[width=0.5\textwidth]{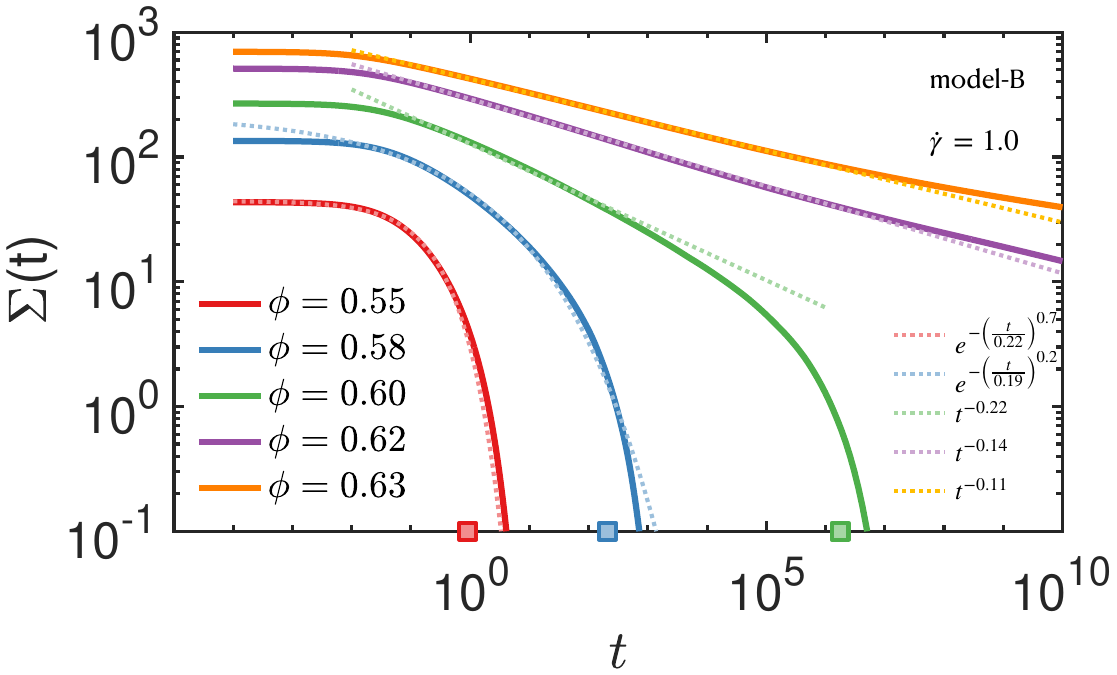}
    \caption{\textbf{Stress relaxation profiles versus time (dimensionless units) for different packing fractions based on the alternative Model-B introduced in the main text. The results are qualitatively the same as those for Model-A presented in the main text.} Quantitatively, the stress relaxation in Model-B is slower than in Model-A, as physically expected.}
    \label{fig.S1}
\end{figure}
\begin{figure*}[h]
    \centering
    \includegraphics[width=1.0\textwidth]{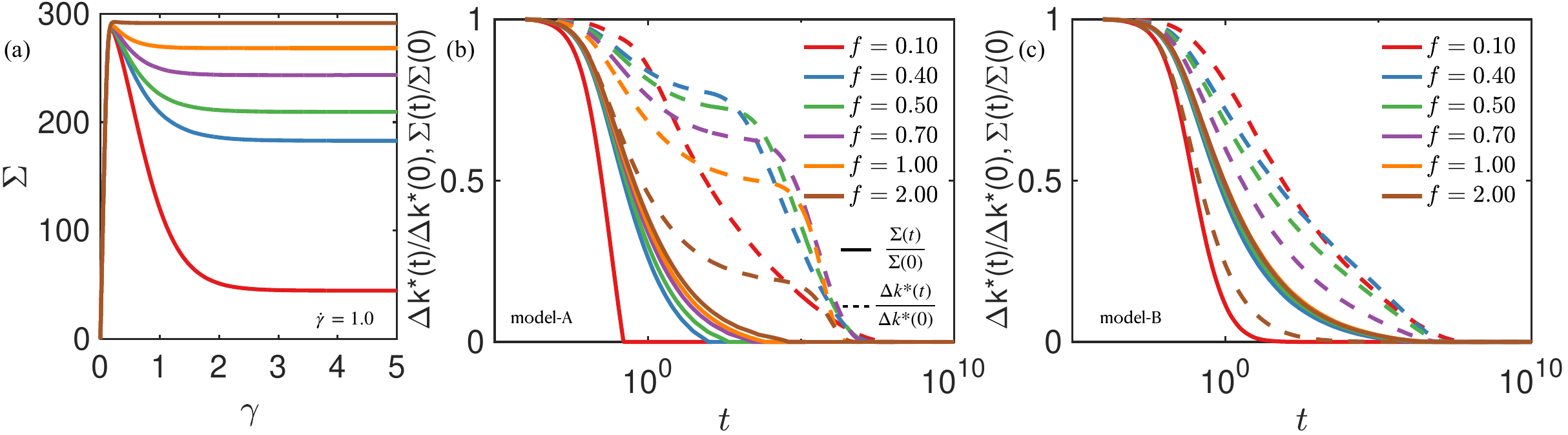}
    \caption{\textbf{Role of stress and structural relaxation times mismatch on the relaxation profiles:} Panel (a) shows the start-up shear response with different mismatch factors, $f$, in Eq(\ref{eqn:3}) that lie in the range of $0.1$ to $2.0$ as color coded and defined in the legends of panels (b) and (c). For ${f}={1}$ as studied in the main article, at the level of the elementary driving force in the evolution equations for structural and stress relaxation the relevant rate is the same and controlled by ${\tau}_{\alpha}$. For ${f}<{1}$, structural relaxation is slower than stress relaxation by a factor of ${f}$, and vice versa for ${f}>{1}$. As discussed in the main text, we believe the scenario with ${f}<{1}$ is more physically realistic. This mismatch, with slower structural relaxation for ${f}<{1}$, leads to an increased overshoot in the start-up shear stress-strain profile. Panels (b) and (c) present the stress and structural relaxation profiles following shear cessation from the steady states shown in panel (a) for Model-A and Model-B, respectively, scaled by their respective steady-state values. For Model-B (panel (c)), the response is relatively simple: stress relaxation profiles are minimally affected by changes in ${f}$, while structural relaxation becomes slower as ${f}$ decreases. In contrast, the Model-A (panel (b)) results show a more significant effect of the mismatch factor. As ${f}$ decreases, the plateau value of structural relaxation increases, which in results in slower stress relaxation. Interestingly, a universal behavior emerges: the structural relaxation time (taken as the time to reach approximately $90\%$ of the quiescent value) remains relatively constant with varying ${f}$, reinforcing the inference made in the main text that structural relaxation is slower than stress relaxation. Overall, the core results presented in the main text remain qualitatively unchanged with variations in ${f}$.}
    \label{fig.S2}
\end{figure*}
\begin{figure}
    \centering
    \includegraphics[width=0.5\textwidth]{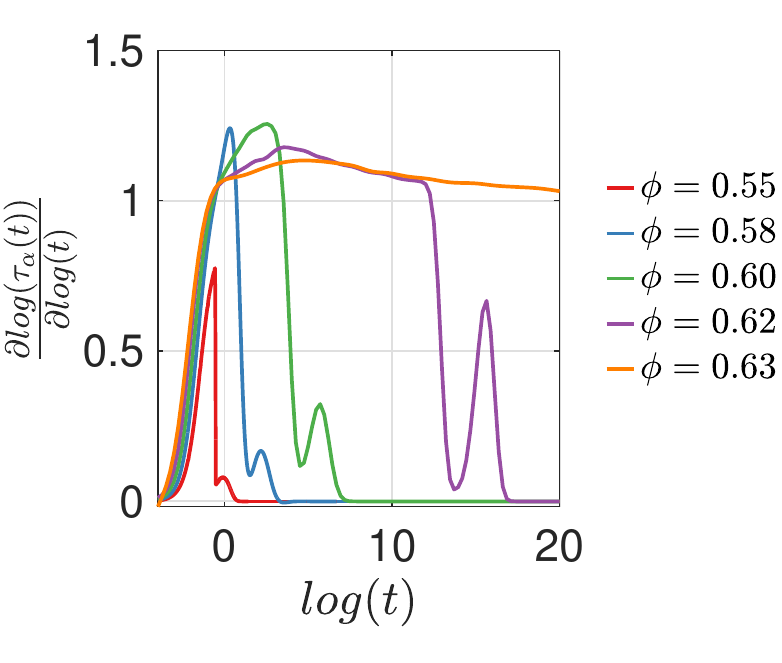}
    \caption{\textbf{Evolution of the instantaneous time scaling exponent characterizing the apparent power-law aging behavior in Fig.\ref{fig3} of the main text.} The apparent time-evolving exponent defined along the y-axis as a logarithmic derivative begins at zero and then rapidly increases to a value slightly exceeding the ideal aging exponent of unity, before gradually converging toward unity at long times. The secondary peak reflects a final stage of aging driven by slow structural relaxation that occurs after the stress has already decayed in Model-A.}
    \label{fig.S3}
\end{figure}
\end{document}